\documentclass{aa}  
\usepackage{graphicx}
\usepackage{txfonts}
\def\ltapprox{\raise 2pt \hbox {$<$} \kern-1.1em \lower 5pt \hbox {$\approx$}}

\def\ltsim{\; \raise0.3ex\hbox{$<$\kern-0.75em \raise-1.1ex\hbox{$\sim$}}\; }
\def\gtsim{\; \raise0.3ex\hbox{$>$\kern-0.75em \raise-1.1ex\hbox{$\sim$}}\; }
\def\arcsec{$^{\prime\prime}\,$}

\def\ie{{\it i.e.,~}}

\def\eg{{\it e.g.,~}}

\begin{document}
   \title{Radio Halos in future surveys in the radio continuum}


   \author{R. Cassano\inst{1}\fnmsep\thanks{\email{rcassano@ira.inaf.it}}, G. Brunetti\inst{1}, Ray P. Norris\inst{3}, H. J. A. R\"ottgering\inst{4}, M. Johnston-Hollitt\inst{2}, M. Trasatti \inst{5}
   } 
\authorrunning{R. Cassano et al.}

   \offprints{R.Cassano}

   \institute{INAF - Istituto di Radioastronomia, via P. Gobetti 101,I-40129 Bologna, Italy\\
   \and School of Chemical \& Physical Sciences, Victoria University of Wellington, Wellington 6140, New Zealand \\
   \and CSIRO Astronomy \& Space Science, Epping, NSW 1710, Australia \\
   \and Leiden Observatory, Leiden University, Oort Gebouw, P.O. Box 9513, 2300 RA Leiden, The Netherlands \\
  \and Argelander-Institut f\"ur Astronomie, Auf dem H\"ugel 71, 53121 Bonn, Germany \\
}


\abstract
{} 
{Giant radio halos are Mpc-scale synchrotron sources 
detected in a significant fraction of massive and merging galaxy clusters.
The statistical properties of radio halos can be used to discriminate 
among various models for the origin of non-thermal particles in galaxy clusters. 
Therefore, theoretical predictions are important as new radio telescopes are about to begin to 
survey the sky at low and high frequencies with unprecedented sensitivity.}
{We carry out Monte Carlo simulations to model the formation and evolution 
of radio halos in a cosmological framework and extend previous calculations 
based on the hypothesis of turbulent-acceleration. 
We adopt a phenomenological approach by assuming that radio halos are either generated in turbulent
merging clusters, or are purely hadronic sources generated in more relaxed clusters,
``off-state'' halos.}
{The models predict that the luminosity function of radio halos at high radio luminosities
is dominated by the contribution of halos generated in turbulent 
clusters. The generation of these halos becomes less efficient
in less massive systems causing a flattening of the luminosity function
at lower radio luminosities, as also pointed out in previous studies.
However, we find that potentially this can be more than compensated for
by the intervening contribution of ``off-state'' halos that dominate at lower radio luminosities.
We derive the expected number of halos to explore the potential of the EMU+WODAN surveys 
that will be carried out with ASKAP and Aperitif, respectively,  in the near future. 
By restricting to clusters at redshifts $\leq$ 0.6, we show that 
the planned EMU+WODAN surveys at 1.4 GHz have the potential to detect up to about 200 new
radio halos, increasing their number by one order of magnitude. 
A fraction of these sources will be ``off-state'' halos that should be found at flux level
$f_{1.4} \leq 10$ mJy, presently accessible only to deep pointed observations.
We also explore the synergy between surveys at different radio frequencies,
the Tier 1 LOFAR survey at 150 MHz and the EMU+WODAN surveys at 1.4 GHz.
We predict a larger number of radio halos in the LOFAR survey due to
the high LOFAR sensitivity, but also due to the existence of halos with very
steep spectrum that glow up preferentially at lower frequencies.
These halos are only predicted in the framework 
of turbulent re-acceleration models and should not have counterparts
in the EMU+WODAN surveys, thus the combination of the two
surveys will test theoretical models.}
{}


\maketitle

\section{Introduction}

Radio halos are diffuse Mpc--scale radio sources observed at the 
center of $\sim 30\%$ of massive galaxy clusters 
(\eg Ferrari et al. 2008, Venturi 2011, Feretti et al. 2012 for recent reviews).
These sources emit synchrotron radiation produced by 
GeV electrons diffusing through
$\mu$G magnetic fields and provide the most important evidence 
of non-thermal components in the intra-cluster medium (ICM). 

Clusters hosting radio halos always display evidence of very recent 
or ongoing merger events (\eg Buote 2001; 
Schuecker et al 2001; Govoni et al. 2004; 
Venturi et al. 2008; Cassano et al. 2010a). 
The connection between radio halos 
and cluster mergers suggests that the gravitational process 
of cluster formation provides the energy to generate the non-thermal 
components in clusters through the acceleration of high-energy particles 
via shocks and turbulence (e.g., Sarazin 2004, Brunetti 2011a, for review).
A scenario proposed to explain the origin of the synchrotron emitting 
electrons in radio halos assumes that seed relativistic electrons 
in the ICM are 
re-accelerated by the interaction with merger-driven MHD turbulence 
in merging galaxy clusters ({\it turbulent re-acceleration} model,
\eg Brunetti et al. 2001; Petrosian 2001). 
According to this scenario, the formation and evolution of radio halos 
are tightly connected to the dynamics and 
evolution of the hosting clusters. 
The occurrence of radio halos at any redshift depends on the rate 
of cluster-cluster mergers and on the fraction of the merger 
energy channelled into MHD turbulence and re-acceleration 
of high energy particles. This latter point depends on microphysics 
of the ICM and it is difficult 
to estimate (see \eg Brunetti \& Lazarian 2011a). 

Despite model details, one of the most important expectations 
of this scenario is the existence of a population of radio halos 
with very steep radio spectra which will be mostly visible 
al low radio frequencies (Cassano et al. 2006; Brunetti et al. 2008). 
The existence of this population can be tested with the upcoming 
surveys at low radio frequencies with the 
Low Frequency Array (LOFAR)\footnote{http://www.lofar.org},
the Long Wavelength Array (LWA, \eg Ellingson et al. 2009) and the 
Murchison Widefield Array (MWA, \eg Tingay et al. 2012). 
Cassano et al. (2010b) have shown that the 
LOFAR {\it Tier-1} ``Large Area Survey'' at 120-190 MHz 
(R\"ottgering 2010), with an expected rms sensitivity of 
$\sim$0.1-0.3 mJy/beam, should detect about 350 giant radio halos 
up to redshift $z\sim0.8$, and half of them should have 
very steep radio spectra (with $\alpha\gtsim1.9$, $F(\nu)\propto \nu^{-\alpha}$).

An alternative scenario relies on the production of secondary
electrons. Inelastic collisions between cosmic ray protons and target thermal 
protons in the ICM produce a population of secondary electrons 
that can generate diffuse synchrotron emission on galaxy 
cluster-scale (\eg Dennison 1980; Blasi \& Colafrancesco 1999). 
Despite several observations pointed out that pure secondary models have 
difficulties in explaining the spectral and morphological properties 
of several nearby giant radio halos (\eg Brunetti et al. 2008; Donnert et al. 2010; 
Brown \& Rudnick 2011; Jeltema \& Profumo 2011, for additional constraints based on
$\gamma$-ray upper limits), synchrotron 
emission produced by secondary electrons must be present in galaxy clusters.
This comes from the theoretical argument that galaxy clusters are
efficient reservoirs of cosmic ray protons (accelerated by structure formation shock waves, 
injected from radio galaxies, or from supernova driven galactic winds) and consequently these protons 
accumulate with cluster life-time increasing the probability
to have proton-proton collisions (V\"olk et al 1996; Berezinsky et al 1997; En\ss lin et al 1997).

Recent Fermi-LAT and Cherenkov-telescopes observations 
have placed upper limits on the ratio of non-thermal CRp to thermal 
energy densities at the level of $\sim$ few\% in a number of
nearby clusters (Ackermann et al. 2010; Aleksi{\'c} et al. 2012).
Assuming these constraints, it has been suggested
that during cluster mergers the re-acceleration 
of secondary electrons by compressible MHD turbulence can generate synchrotron radiation in good agreement 
with radio halos, while synchrotron emission $\sim 10$ times fainter 
is generated in dynamically relaxed clusters when turbulence is dissipated (Brunetti \& Lazarian 2011b).  
This theoretical conjecture is consistent with the observed radio 
bimodality in galaxy clusters (\eg Brunetti et al. 2009; 
Cassano et al. 2010a), although it predicts a level of emission 
(from pure secondaries) in relaxed clusters that is close the upper limits 
derived for these clusters from present radio observations.
More recently, Brown et al. (2011) claimed the detection of radio emission from 
the stacking of SUMMS images of  ``off-state'' (non-radio halo) 
clusters at a level $\sim 10$ times fainter than that of
classical radio halos, potentially in line with the above theoretical picture.

In about the next 10 years several revolutionary radio telescopes 
will survey the sky with unprecedented 
sensitivity and spatial resolution at very low (LOFAR, LWA, MWA) and
GHz frequencies (ASKAP; Johnston et al. 2008 and 
Aperitif; Oosterloo, Verheijen, van Cappellen 2010).
This gives the opportunity to constrain the complex connection 
between cluster dynamics and diffuse radio emission in galaxy clusters by 
means of statistical studies of adequately large cluster sample. 
The combination of incoming surveys at low (LOFAR) and higher
(e.g., the EMU survey with ASKAP) 
frequencies will add considerable value in the
attempt to discriminate between different physical origins of
giant radio halos in galaxy clusters and, in general, of
non-thermal cluster components.
The first LOFAR observation at $\sim 63$ MHz of the giant radio 
halo in Abell 2256 shows that the radio spectrum extracted in the region of the halo steepens at lower frequencies. 
This unexpected result suggests that physical scenario more complex than previously thought should be considered 
to explain the formation of the radio halo in this cluster (van Weeren et al. 2012). In fact, different populations of relativistic electrons
may coexist in the volume of the radio halo if they originate from different acceleration mechanisms,
or if electrons are accelerated in a non homogeneous turbulent region where the efficiency of particle acceleration change 
with space and time in the emitting volume (van Weeren et al. 2012).

In this paper, we extend previous statistical modeling of giant radio halos by combining a picture based 
on turbulent re-acceleration of relativistic electrons by MHD turbulence in cluster mergers with
the process of continuous injection of secondary electrons via p-p collision in the ICM. 
This provide a novel approach to interpret future data from surveys of galaxy clusters.
This has not the aim to reproduce particular spectral features such as those observed in Abell 2256, 
but its aim is to provide a simplified (but viable) description of the general observed properties of the population
of radio halos in galaxy clusters.

In Sect. 2 we summarize the main ingredients used in the model calculations, 
derive the occurrence of radio halos in clusters (Sect. 2.2) and 
the expected radio halo luminosity functions (Sect. 2.3). 
In Sect. 3 we discuss and model
the contribution to the diffuse cluster-scale
synchrotron emission from secondary electrons.
In Sect. 4 we describe the EMU and WODAN surveys, while in Sect. 5 we
derive the expected number of 
radio halos at 1.4 GHz and discuss the potential of the EMU and WODAN surveys. 
In Sect. 6 we discuss the potential of combining LOFAR and EMU and WODAN
surveys in addressing the physics of giant radio halos. 
Our conclusions are given in Sect.7.

\noindent
A $\Lambda$CDM ($H_{o}=70$ $\mathrm{ Km\, s^{-1} Mpc^{-1}}$, $\Omega_{m}=0.3$, 
$\Omega_{\Lambda}=0.7$) cosmology is adopted throughout the paper.

\section{Statistical modelling of giant radio halos from 
turbulent re-acceleration}

\subsection{Main ingredients}
\label{steps}

\begin{figure}
\begin{center}
\includegraphics[width=0.5\textwidth]{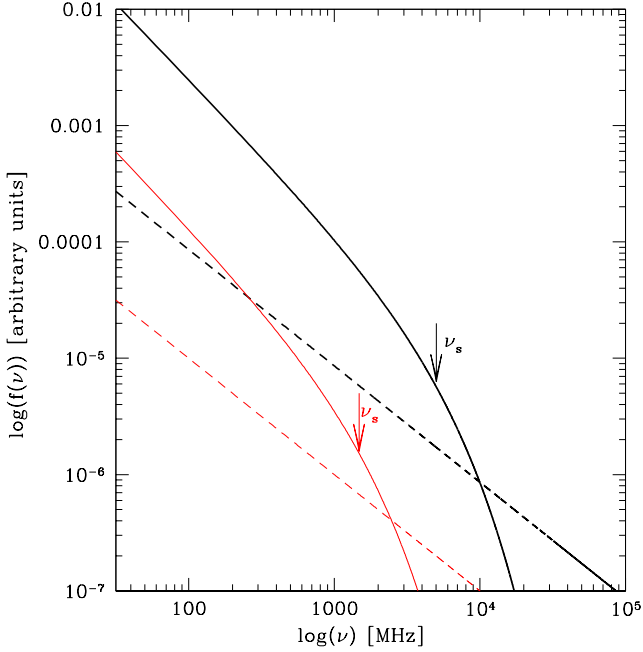}
\caption[]{Reference spectra of ``turbulent'' radio halos (solid lines) and ``off-state'' (hadronic) halos (dashed lines) in a massive 
(\ie $M_v\sim 2.5\times 10^{15}\, M_{\odot}$; black lines) and less massive (\ie $M_v\sim 10^{15}\, M_{\odot}$; red lines) cluster. 
Arrows indicate the position of the steepening frequency, $\nu_s$, in the two cases. The turbulent spectra are computed assuming
in both cases a merger event with a sub-clump of mass $\Delta M=5\times 10^{14}\,M_{\odot}$ at $z=0.023$.}
\label{Fig.spettri}
\end{center}
\end{figure}

Turbulence generated during cluster mergers may re-accelerate 
relativistic particles and produce diffuse synchrotron emission 
from Mpc regions in galaxy clusters (\eg Brunetti et al. 2008). 
Diffuse radio emission in the form of giant radio halos should 
be generated in connection with massive mergers and fade away as 
soon as turbulence is dissipated and the emitting electrons cool 
due to radiative losses. 
Attempts for estimating the statistical properties of giant radio halos 
in the context of this scenario have been carried out in the 
past few years (Cassano \& Brunetti 2005; Cassano et al. 2006; 
Cassano et al. 2010b). These studies allowed for a comparison 
with the presently observed statistical properties of giant radio halos, 
while at the same time they provide predictions to test with future instruments. 

In these calculations we
model the properties of the halos and their cosmic evolution by means of 
a Monte Carlo approach, taking into account the main processes 
that play a role in this scenario: the rate of cluster-cluster mergers 
in the Universe and their mass ratios, and the fraction of the energy 
dissipated during these mergers that is channelled into MHD turbulence 
and acceleration of high energy particles. 
We refer the reader to the papers quoted above for details, 
here for the sake of completeness we briefly report the essential steps of those calculations : 

\begin{itemize}
\item[{\it i)}]
The formation and evolution of dark matter halos of galaxy clusters is computed 
by the extended Press \& Schechter approach (1974, hearafter PS; 
Lacey \& Cole 1993), which is based on the hierarchical theory 
of cluster formation. 
Given the present-day mass and temperature of the parent clusters, 
the cluster merger history ({\it merger trees}) is obtained 
by using Monte Carlo simulations. 

\item[{\it ii)}]
The generation of the turbulence in the ICM is estimated 
for each merger identified in the {\it merger trees}. Turbulence 
is assumed to be generated (and dissipated) within a timescale 
of the order of the cluster-cluster crossing time in that merger. 
Furthermore, it is assumed that the turbulence is generated in 
the volume swept by the subcluster infalling into the main cluster.
The injection rate of turbulent fast modes$/$waves, that are used to 
calculate particle acceleration, is assumed to be a fraction, $\eta_t$, 
of the $PdV$ work done by this subcluster.

\item[{\it iii)}]
The resulting spectrum of MHD turbulence generated by the chain of 
mergers in any synthetic cluster and its evolution with cosmic time 
is computed by taking into account the injection of waves 
and their damping, due to thermal and relativistic particles, 
in a collisionless plasma. 
Acceleration of particles by this turbulence and their evolution is 
computed in connection with the evolution of synthetic clusters 
by solving Fokker-Planck equations and including all the relevant 
energy losses of particles.

\item[{\it iv)}]
Synchrotron losses, particle acceleration and emissivity are calculated assuming homogeneous
models (see Cassano et al. 2006): radio halos are assumed to be homogeneous spheres of radius $R_H\sim 500\,h_{50}^{-1}$ kpc,
and volume-average values of turbulent energy, acceleration rate and magnetic field are adopted.
We assume a value of the magnetic field which scales with the virial 
mass of clusters, $M_v$ as

\begin{equation}
<B> \, = B_{<M>} \big({{M_v}\over{<M>}} \big)^{b}\,,
\label{b}
\end{equation}

\noindent
where $b>0$, and $B_{<M>}$ is the value of the rms magnetic field associated with a cluster with
mass $<M>\simeq 1.6\times 10^{15}\,M_{\odot}$ (see next
Sect. and Sect.~3 in Cassano et al. 2006, for details). This scaling is motivated by the results of cosmological MHD simulations
that found that the magnetic field scales with the temperature (and mass) of the simulated clusters
(\eg Dolag et al. 2002)\footnote{Dolag et al. (2002) found a scaling $B\propto T^2$, which would imply
that $B\propto M^{1.33}$ if the virial scaling $M\propto T^{3/2}$ is assumed.}.

\end{itemize}

\subsection{Occurrence of Radio Halos with redshift and cluster mass}

Stochastic particle acceleration by MHD turbulence is a rather inefficient 
process in the ICM that accelerates electrons 
at energies $m_e c^2 \gamma_{max} \leq$ several GeV, 
since at higher energies the radiation losses quench 
the acceleration process (see e.g., Brunetti 2011b and references therein). 
In the case of a homogeneous model this implies a steepening of the synchrotron spectrum that affects 
the capability to detect radio halos at frequencies substantially larger 
than the frequency, $\nu_s$, where the steepening becomes severe. 
The frequency $\nu_s$ depends on the acceleration efficiency , $\chi$, 
and on $<B>$,  as $\nu_s\propto <B>\chi^2 /(<B>^2+B_{cmb}^2)^2$ 
(\eg Cassano et al.~2006, 2010b). Monte Carlo simulations of cluster 
mergers that occur during the hierarchical process of cluster 
formation (Sect.~2.1) allow for evaluating $\chi$ from the estimated 
rate of turbulence-generation and the physical condition in the ICM, 
and consequently to explore the dependence of $\nu_s$ on cluster 
mass, redshift, and merger parameters in a statistical sample of 
synthetic clusters.

\begin{figure*}
\begin{center}
\includegraphics[width=0.45\textwidth]{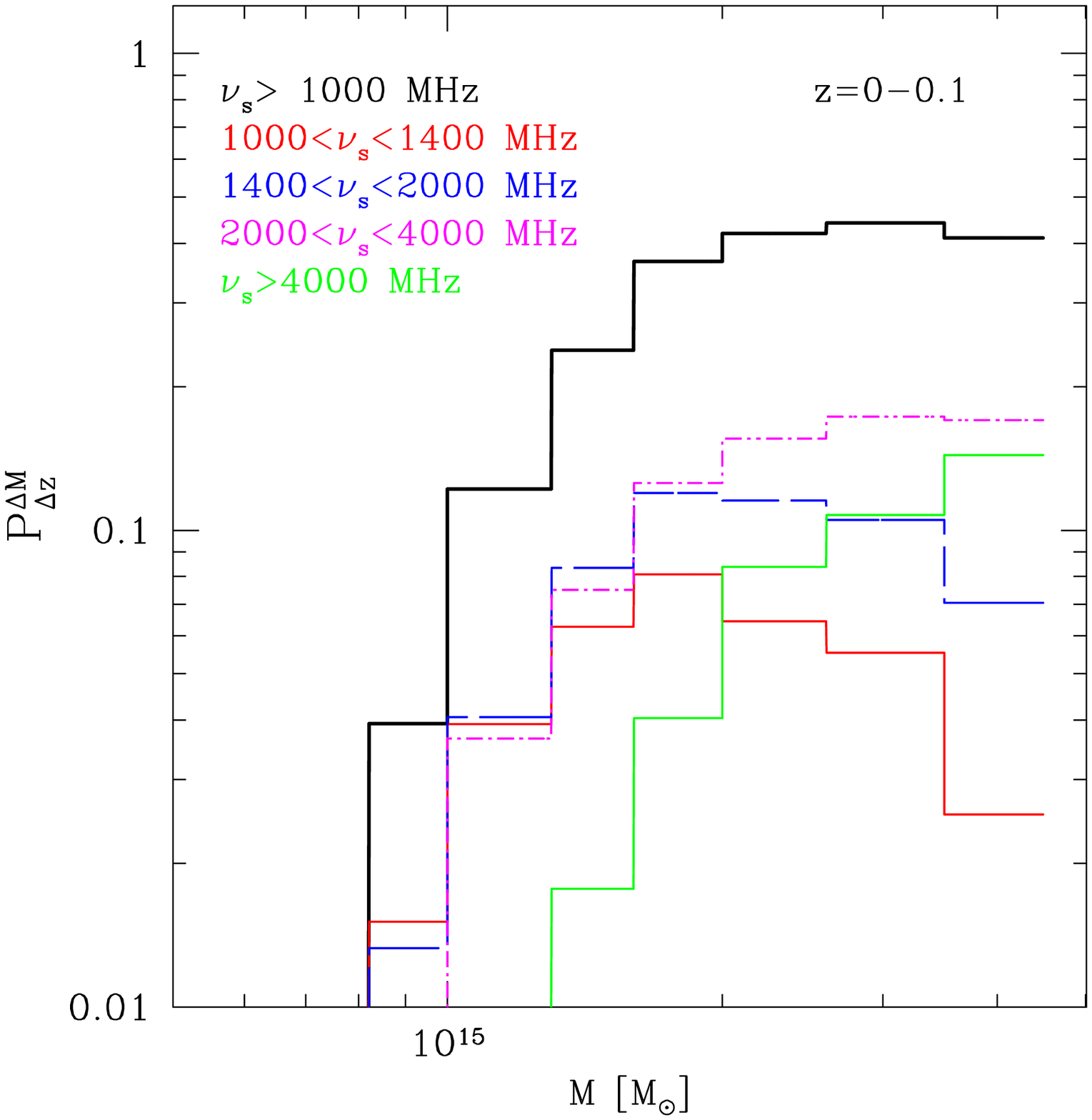}
\includegraphics[width=0.45\textwidth]{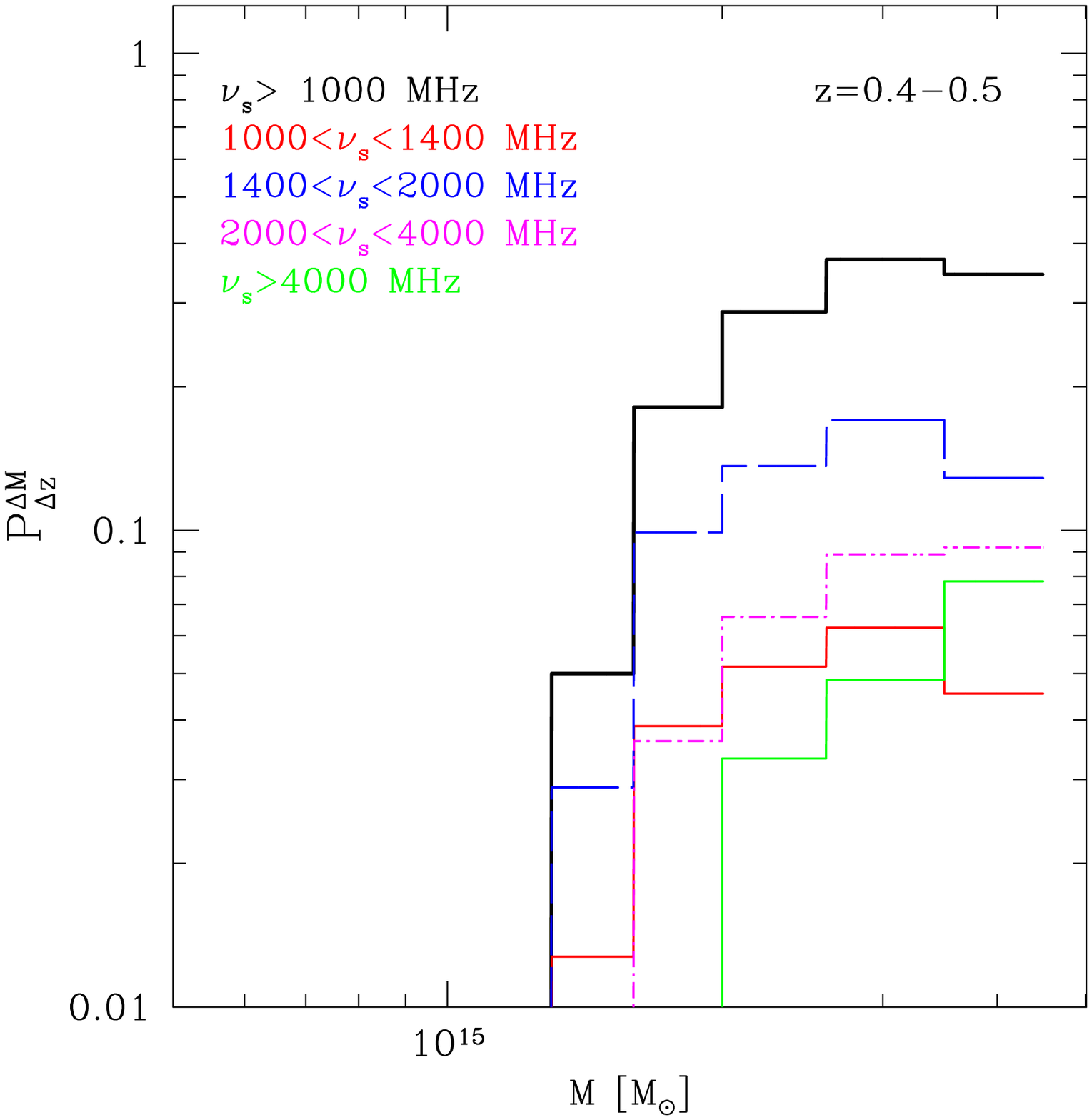}
\caption[]{Fraction of clusters with radio halos with $\nu_s \geq$ 1000 MHz (black, upper, solid lines) as a function of
the cluster mass in the redshift range $0-0.1$ (left panel) and $0.4-0.5$ (right panel). 
The fractions of clusters with radio halos with $\nu_s$ in different frequency ranges are also shown : 
$1000<\nu_s < 1400$ MHz (red lines), $1400<\nu_s <2000$ MHz (blue lines), $2000<\nu_s<4000$ MHz (magenta lines),
and $\nu>4000$ MHz (green lines).}
\label{Fig.fRH}
\end{center}
\end{figure*}

\noindent A reference example of spectra of ``turbulent'' radio halos as modeled
in the present paper is shown in Fig.\ref{Fig.spettri}. A simplified approach to estimate the occurrence of radio halos 
in surveys at different observing frequencies is to assume that 
only those halos with $\nu_s \geq \nu_o$ can be observable, 
$\nu_o$ being the observing frequency. 
Energy arguments imply that giant radio halos with $\nu_s \geq$ 1 GHz 
are generated in connection with the most energetic merger-events 
in the Universe. Only these mergers can generate enough turbulence 
on Mpc scales and give the acceleration rate 
that is necessary to maintain the relativistic electrons emitting at 
these frequencies (Cassano \& Brunetti 2005). 
Present surveys carried out at $\nu_o \sim$ 1 GHz detect radio halos 
only in the most massive and merging clusters (\eg Venturi 2011 and
references therein, for a review).
The fact that $\sim 1/3$ of X-ray luminous galaxy clusters host giant
radio halos observed at $\sim$ GHz frequencies has been used 
to constrain the parameters for the generation of turbulence in 
the modeling described in Sect.~2.1 ($\eta_t \approx 0.1-0.3$, 
Cassano \& Brunetti 2005; Cassano et al 2008a).

In Fig.~\ref{Fig.fRH}, we plot the fraction of radio halos 
with $\nu_s \geq$ 1 GHz (black upper line) and the differential 
contribution to this fraction from radio halos with $\nu_s$ in four 
frequency ranges (see figure caption for details). 
This is obtained by assuming a {\it reference} set of model parameters, 
namely $<B>=1.9\, \mu$G, $b=1.5$, $\eta_t=0.2$ (see also Cassano et al.~2006). 

At low redshift ($z=0-0.1$), the fraction of clusters hosting radio halos 
with $\nu_s \geq$ 2 GHz increases with the cluster mass, reaching 
a maximum value in the case of very massive clusters 
($M_v > 2-3\times10^{15}\,M_{\odot}$). On the other hand, the fraction 
of clusters hosting radio halos with $\nu_s$ in the range 1-2 GHz (red 
and blue lines) reaches a maximum for clusters masses slightly 
below $\sim 2\times10^{15}\,M_{\odot}$, and then decreases for more 
massive systems. This behavior is due to the fact that in general,
in our model, the fraction of clusters with radio halos increases with the cluster 
mass, since more massive clusters are more turbulent, and thus are more likely to host a radio halo.
The occurrence of halos with relatively lower values of $\nu_s$ ($\nu_s< 2$ GHz) also 
increase with the cluster mass, but lower the value of $\nu_s$ lower the 
value of the mass up to which the occurrence of these halos increases, 
for larger masses their occurrence decrease since those clusters preferentially
form halos with larger values of $\nu_s$.

Consequently we expect that very massive (and hot) clusters tend 
to generate giant radio halos with radio spectra flatter (higher 
values of $\nu_s$) than those in less massive systems. 
A tendency to have halos with flatter spectra in more massive (hot)
systems have been reported in literature (Feretti et al. 2004; 
Giovannini et al. 2009, Venturi et al. 2012, sub.).
If confirmed these findings would support our expectations.

In addition, we find that giant radio halos with higher values of $\nu_s$ 
become rarer with increasing redshift, mainly because of
the inverse Compton losses that increase with redshift and
limit the maximum energy of the accelerated electrons. 

\begin{figure}
\begin{center}
\includegraphics[width=0.5\textwidth]{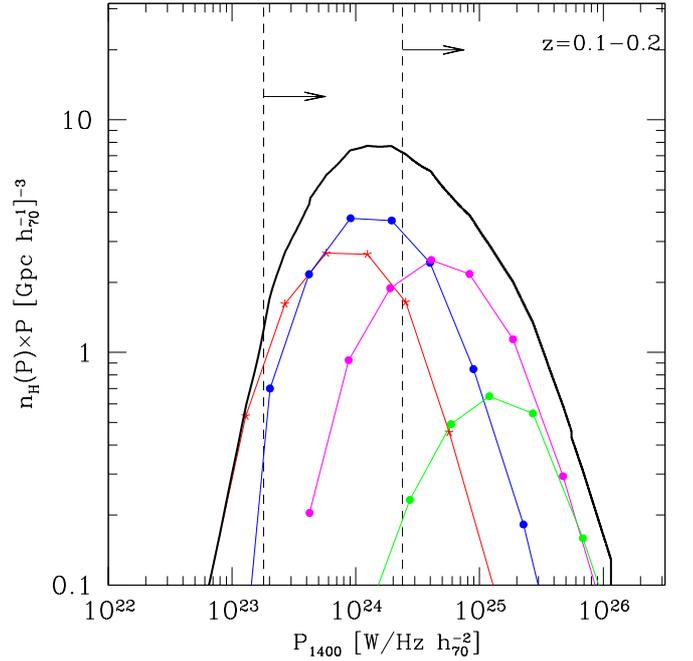}
\caption[]{Radio halo luminosity function at $\nu_o$=1400 MHz (black
lines) for clusters at redshift $0.1-0.2$. 
The contributions from halos with $\nu_s$ in different frequency ranges are also shown : 
$1000<\nu_s < 1400$ MHz (red lines), $1400<\nu_s <2000$ MHz (blue lines), 
$2000<\nu_s<4000$ MHz (magenta lines), and $\nu>4000$ MHz (green lines).
The dashed lines correspond to the minimum detectable radio halo power 
(according to Eq.~\ref{fmin} with $\xi_1=3$, see text for details)
for EMU + WODAN survey and for the NVSS (from left to right).}
\label{Fig.RHLF}
\end{center}
\end{figure} 

\begin{figure}
\begin{center}
\includegraphics[width=0.5\textwidth]{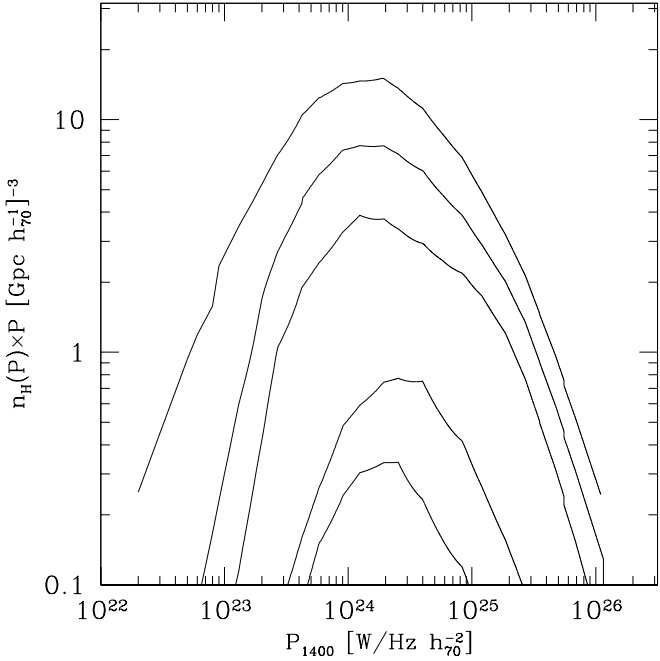}
\caption[]{Radio halo luminosity function at $\nu_o$=1400 MHz derived in the redshift intervals:
z=$0-0.1$, $0.1-0.2$, $0.2-0.3$, $0.3-0.4$, $0.4-0.5$ and $0.5-0.6$ (from top to bottom).
}
\label{Fig.RHLF_z}
\end{center}
\end{figure} 

\subsection{The radio halo luminosity function}
\label{Sect.RHLF}

The luminosity functions 
of radio halos (RHLFs) with $\nu_s\geq \nu_0$ 
(\ie the expected number of halos per comoving volume and radio power ``observable'' at
frequency $\nu_0$) 
can be estimated by :

\begin{equation}
{dN_{H}(z)\over{dV\,dP(\nu_0)}}=
{dN_{H}(z)\over{dM\,dV}}\bigg/ {dP(\nu_0)\over dM}\,,
\label{RHLF}
\end{equation}

\noindent
where $dN_{H}(z)/dM\,dV$ is the theoretical mass function of radio 
halos with $\nu_s \geq \nu_0$, that is obtained by combining Monte Carlo 
calculations of the fraction of clusters with radio halos and 
the PS mass function of clusters (\eg Cassano et al.~2006).

$dP(\nu_0)/dM$ depends on (is proportional to) the unknown number of seeds electrons 
(or that of cosmic ray protons) in the ICM. Following Cassano et al.(2006) we estimate 
$dP(\nu_0)/dM$ from the correlation observed for giant radio halos
between the 1.4 GHz radio power, $P(1.4)$, and the mass of the 
parent clusters (\eg Govoni et al. 2001; Cassano et al. 2006). 
The observed correlation allows us to normalize the radio
luminosity of giant radio halos under the assumption that
``all'' radio halos with $\nu_s >$ 1 GHz follow the correlation.
In addition, in the context of the modeling described in Sect.~2.1, 
the value of the derivative $dP(1.4)/dM$ depends on the set of parameters
($B_{<M>}$, $b$), consequently the slope of the observed $P(1.4)-M$ 
correlation ($P\propto M^{\alpha_M}$ with $\alpha_M=2.9\pm0.4$) 
can be used to constrain model parameters selecting an allowed region 
in the parameter space ($B_{<M>}$, $b$) (Cassano et al.~2006). 
In this paper we shall adopt a {\it reference} set of model parameters, 
\ie $B_{<M>}=1.9\, \mu$G and $b=1.5$ (and $\eta_t=0.2$), 
that falls in this region; in this case $\alpha_M\approx 3.3$.

To derive the RHLF at frequency $\nu_o=1400$ MHz, we consider 
the contribution of all radio halos with $\nu_s > 1$ GHz.
We first obtained the RHLF for halos with $\nu_s$ in the frequency 
interval ${\Delta \nu_s}_i$, and then combined 
the different contributions from the considered intervals ${\Delta \nu_s}_i$ :

\begin{equation}
{dN_{H}(z)\over{dV\,dP(\nu_o)}}=
\sum_i 
\big( {dN_{H}(z)\over{dM\,dV}} \big)_{{\Delta \nu_s}_i}
\big( {dP(\nu_o)\over dM} \big)_{{\Delta \nu_s}_i}^{-1}\,.
\label{RHLF_nuo}
\end{equation}

\noindent The relation between the monochromatic radio luminosity $P(\nu_o)$ of halos with a given 
$\nu_s$ (with $\nu_s\geq \nu_0$) and that of halos with $\nu_s=1.4$ GHz has been derived by Cassano et al.(2010b) :

\begin{equation}
P_{\nu_s}(\nu_o, M_v)=
P_{\nu_s}(\nu_s, M_v)\Big(\frac{\nu_s}{\nu_o}\Big)^{\alpha}
=
P_{1.4}(1.4, M_v)
\Big(\frac{\nu_s}{\nu_o}\Big)^{\alpha}\, ,
\label{scale2}
\end{equation}

\noindent
allowing a prompt evaluation of $(dP(\nu_o)/dM)_{{\Delta \nu_s}_i}$ in Eq.~\ref{RHLF_nuo}.

\noindent
Taking $P_{1.4}(\nu_0,M_v)=P_{1.4}(1.4,M_v)(1.4\,\mathrm{GHz}/\nu_0)^{\alpha}$, 
from Eq.\ref{scale2} one has:

\begin{equation}
P_{\nu_s}(\nu_o, M_v)=
P_{1.4}(\nu_o, M_v) \Big(\frac{\nu_s}{1400\,\mathrm{MHz}}\Big)^{\alpha}\,,
\label{scale3}
\end{equation}

\noindent
\ie radio halos with synchrotron spectra that steepen at lower 
frequencies will also have smaller monochromatic radio powers 
at the observing frequency $\nu_o$.

\noindent 
As a relevant example, 
in Fig.\ref{Fig.RHLF} we report the expected RHLF at 1.4 GHz 
(black lines) for $z=0.1-0.2$ (solid thick lines), where we also mark 
the relative contributions from 
halos with different values of $\nu_s$ (see caption). 

\noindent
As already discussed in Cassano et al.~(2006) and Cassano et al.~(2010b), 
the shape of the RHLF flattens at lower radio powers because of 
the expected decrease of the efficiency of particle acceleration in the case
of less massive clusters. 
We note that halos with $\nu_s > 2$ GHz (green and magenta lines, 
Fig.~\ref{Fig.RHLF}) do not contribute significantly to 
the RHLF at lower radio luminosities. 
This is because higher-frequency halos are generated in very energetic 
merger events that are extremely rare in smaller systems, and have 
large monochromatic radio luminosities (Eq.\ref{scale3}) (red and blue 
lines, Fig.~\ref{Fig.RHLF}).
Finally, we note that the normalization of the
RHLFs decreases with increasing 
redshift (Fig.\ref{Fig.RHLF_z}) due to the evolution with $z$ of both 
the cluster mass function and the fraction of galaxy clusters 
with radio halos (Fig.~\ref{Fig.fRH}, see also Cassano et al.~2006). 

\section{Secondary electrons}
\label{secondary}

An additional contribution to the Mpc-scale 
synchrotron emission from galaxy clusters comes from the process 
of continuous injection of secondary electrons via p-p collisions 
in the ICM. 

\noindent
A self-consistent modeling of the re-acceleration of primary cosmic rays and
secondary electrons$/$positrons by compressible MHD turbulence in
the ICM has been developed in Brunetti \& Lazarian (2011b).
Implementing their formalism in our cosmological Monte Carlo framework
is challenging and out of the main focus of the present paper.

\begin{figure}
\begin{center}
\includegraphics[width=0.4\textwidth]{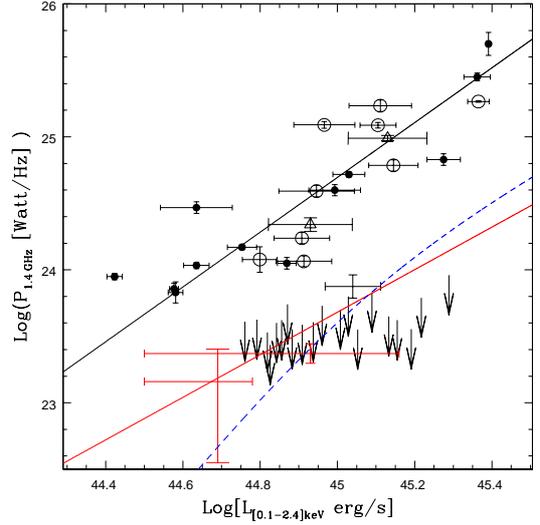}
\caption[]{Correlation between the radio halo luminosity at 1.4 GHz and the cluster X-ray luminosity. Clusters from the literature (filled circle) and clusters from the GMRT RH Survey (Venturi et al. 2008; open circles and black arrows) are reported. The red crosses are obtained by staking the radio images of clusters from the SUMSS survey (Brown et al. 2011). On the same figure we also report the two scalings adopted in the present paper for halos produced by secondary electrons: case a) (blue line) and case b) (red line)(see text for details).}
\label{Fig.Lx_P_sec}
\end{center}
\end{figure}

\noindent
Here we adopt a simplified approach based on two separate
cluster radio-populations.
We assume that presently observed radio halos are mainly generated
in merging clusters by particle acceleration by turbulence.
In more relaxed clusters, turbulence cannot maintain a population
of relativistic electrons emitting at the observing frequencies, $\nu_o$,
and in these cases the dominant contribution to Mpc-scale radio
emission is due to the generation of secondary particles.
As a first approximation we can assume that the level of the
emission from secondary particles in these clusters is stationary, \ie it
does not depend on cluster dynamics, 
since the primary protons accumulate in galaxy clusters over the cluster
lifetime and continuously generate secondaries (Blasi 2001)
\footnote{We do not consider here possible modifications of the synchrotron emission
in these clusters due to magnetic field amplification and cosmic-ray diffusion (\eg Kushnir et al. 2009;  En\ss lin et al. 2011).}.
Under our hypothesis the emission produced by secondary particles 
can be constrained from the limits derived from deep radio 
observations of ``radio quiet'' galaxy clusters 
(\eg Brunetti et al. 2007, 2009). 
These limits are about one order of magnitude below the 
radio-X-ray luminosity correlation for classical giant radio
halos (Fig.~\ref{Fig.Lx_P_sec}). 
More recently, Brown et al. (2011) have detected diffuse emission 
from ``off-state'' galaxy clusters by stacking SUMSS 
images of $\sim$ 100 clusters. 
Potentially, this signal can be contributed by secondary particles 
in less turbulent systems. 
Motivated by these recent observations, here we assume that clusters where turbulence
is not enough to produce giant radio halos emitting at the observing frequency $\nu_0$
(\ie with $\nu_s<\nu_0$, see Sect.2) host diffuse radio emission powered by pure secondaries 
with a luminosity that is similar to the upper limits in Fig.~\ref{Fig.Lx_P_sec}. 
Following Brown et al. (2011) we refer to these halos 
as to ``off-state'' radio halos. 

\noindent The massfunction of  ``off-state'' halos is given by:

\begin{equation}
{{dN_H^{sec}(z,\nu_o)} \over {dV\,dM}}=(1-f_{RH}(M,\nu_o)) 
\times {{dN_M^{cl}}\over {dV\,dM}}
\label{RHMF_sec}
\end{equation}

\noindent where $f_{RH}(M, \nu_o)$ 
is the fraction of clusters of mass $M$ with radio halos 
due to turbulence re-acceleration (with $\nu_s \geq \nu_o$)
and $dN_M^{cl}/dV\,dM$ is the cluster mass function. 
The luminosity function of ``off-state'' halos is:

\begin{equation}
{{dN_H^{sec}(z,\nu_o)} \over {dV\,dP}}=
{{dN_H^{sec}(z,\nu_o)} \over {dV\,dM}}\times {{dM}\over{dP}}\,,
\label{RHLF_sec}
\end{equation}

\noindent 
We derive $dM/dP$ from the expected 
relation between the radio luminosity of ``off-state'' halos 
and the mass (or $L_X$) of the host clusters. 
In their most simple formulation, secondary models predict a correlation 
between the radio luminosity of halos and the cluster X-ray 
luminosity (\eg Kushnir, Katz \& Waxman 2009)\footnote{Here we
modify Kushnir et al. formalism to account also for the case of
weak magnetic fields.} that is slightly flatter than that of 
giant radio halos (see Brunetti et al. 2009):

\begin{equation}
P=A_{norm}^{X}L_X^{1.6}
\times {{B^2}\over{B^2+B_{cmb}^2}}\approx 
A_{norm}^{M}\,M^{2.35}\times {{B^2}\over{B^2+B_{cmb}^2}}\
\end{equation}

\noindent 
where the second equivalence is obtained by assuming 
the $M_v-L_X$ correlation (taken from Cassano et al. 2006). 
The normalization factors $A_{norm}^X$ and $A_{norm}^M$ are 
derived in order to be consistent with the radio upper limits in the plane 
$P_{1.4}-L_X$ obtained for ``radio quiet'' systems 
(Brunetti et al. 2007, 2009).
Specifically, we adopt two different approaches and obtain two 
scalings (see Fig.\ref{Fig.Lx_P_sec}): 
a) in the first one we use the scaling of the magnetic field 
with the cluster mass in Eq.\ref{b} 
with $b=1.5$ and $B_{<M>}=1.9\,\mu$G, and adopt 
$P\approx 4\times 10^{23}$ Watt/Hz for $L_{X} = 10^{45}$ erg/sec 
(blue line in Fig.\ref{Fig.Lx_P_sec}); 
b) in the second one we consider a constant magnetic field $B=3\mu$G 
and a slightly higher normalization, 
$P\approx 5\times 10^{23}$ Watt/Hz for $L_{X} = 10^{45}$ erg/sec 
(red line in Fig.\ref{Fig.Lx_P_sec}). 
The latter approach maximizes the contribution from secondary electrons. A reference example
of spectra of ``off-state'' (hadronic) halos in the case a) is show in Fig.\ref{Fig.spettri}.

\noindent Apparently, the adopted scalings for secondaries are not fully consistent with
non-detections of higher luminosity clusters, however some of these clusters 
are cool-core systems and their luminosity is likely to be boosted (up to $\sim 2$ times)
by the emission produced in the core (Cassano et al. in, prep).
In addition, upper limits reported in Fig.\ref{Fig.Lx_P_sec} have been obtained at 610 MHz and
scaled to 1.4 GHz by adopting a spectral index $\alpha=1.3$, while the spectrum 
of secondary halos assumed in the present paper is $\alpha=1.0$ implying that a fair comparison
with theoretical models should use shallower limits.

\begin{figure*}
\begin{center}
\includegraphics[width=0.4\textwidth]{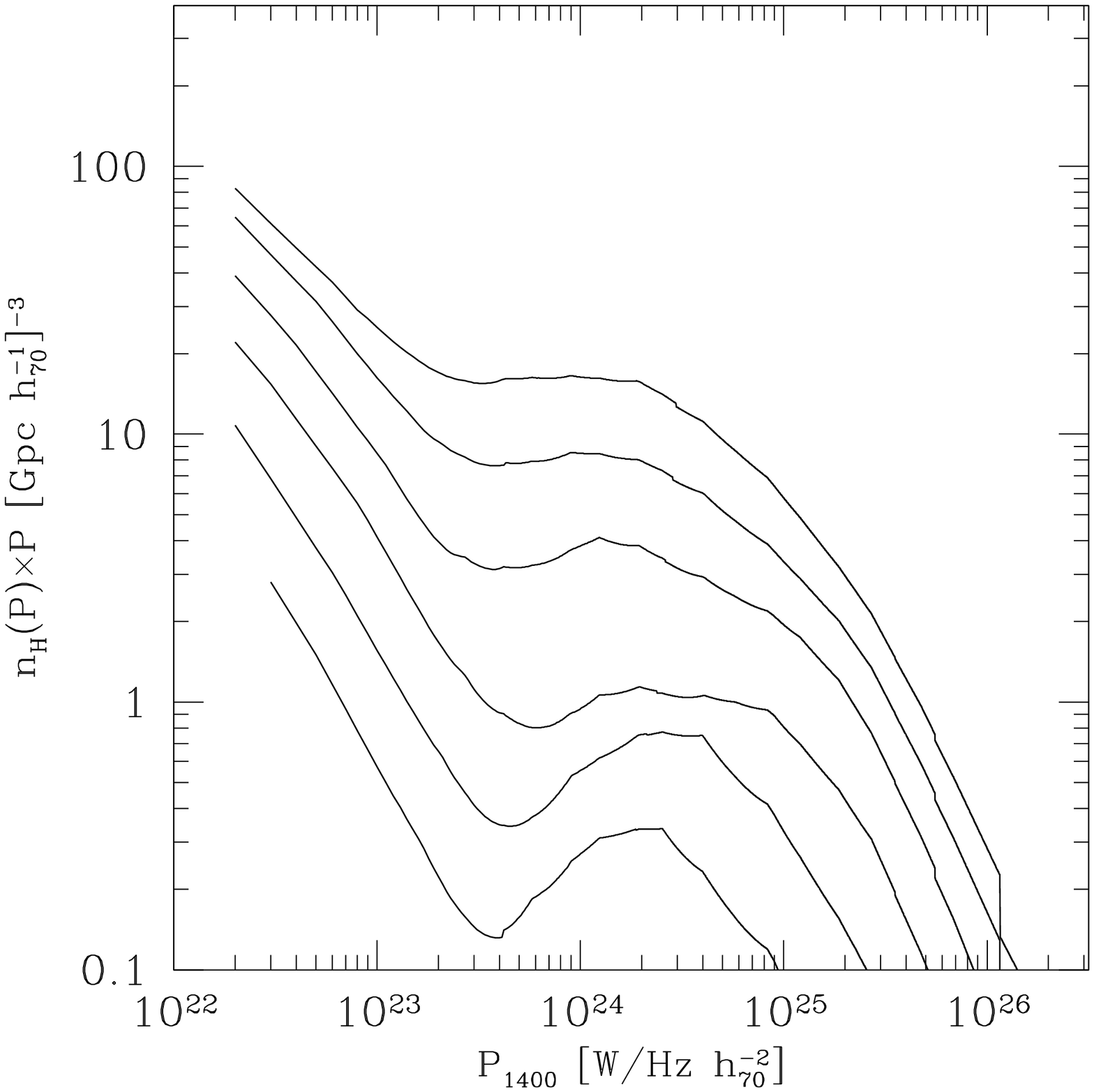}	
\includegraphics[width=0.4\textwidth]{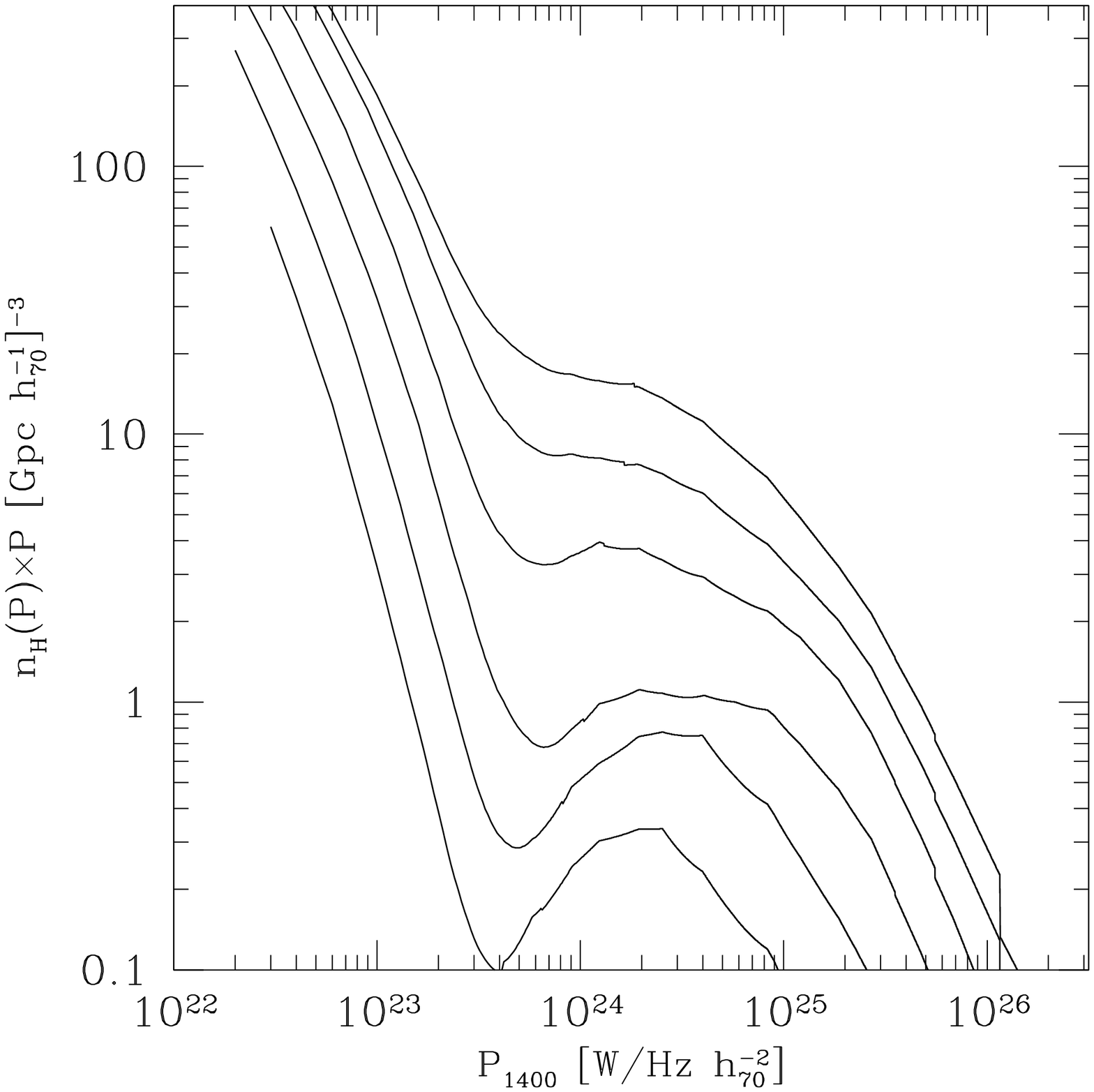}
\caption[]{Total RHLFs obtained by combining the contributions from ``turbulent'' radio halos and from (purely hadronic) ``off-state'' halos in the case a) ({\it left panel}) and b) ({\it right panel}) (see text for details). The RHLF are reported in different redshift interval: z=$0-0.1$, $0.1-0.2$, $0.2-0.3$, $0.3-0.4$, $0.4-0.5$ and $0.5-0.6$ (from top to bottom).}
\label{Fig.RHLF_sec}
\end{center}
\end{figure*}

In Fig.~\ref{Fig.RHLF_sec} we report the total RHLF, obtained 
by combining the contributions from ``turbulent'' radio halos and
from (purely hadronic) ``off-state'' halos, 
in the case a) (left panel) and b) (right panel). 
Under our assumptions, 
``off-state'' halos dominate the RHLF at lower radio luminosities
where the RHLF due to turbulent radio halos flattens.

\section{The EMU+WODAN Surveys}

\subsection{ASKAP: EMU survey}
The Australian SKA Pathfinder (ASKAP) (Johnston et al. 2008) 
is a new radio telescope being built at the Murchison Radio-astronomy Observatory  in Western Australia. It consists of 36 12-metre antennas distributed over a region 6 km in diameter. 
ASKAP will have an instantaneous  field of view of 30 deg$^2$, enabling surveys of a scope that cannot be contemplated with current-generation telescopes. 
The ASKAP array configuration  
balances the need for high sensitivity to extended structures with the need for high resolution. To achieve this, 30 antennas follow a roughly Gaussian distribution with a scale of $\sim$\,700 m, corresponding to a point spread function of $\sim$ 30\arcsec, with a further six antennas extending to a maximum baseline of 6 km, corresponding to a point spread function of $\sim$ 10\arcsec. 
These short spacings of ASKAP deliver excellent sensitivity to low-surface brightness emission, which is essential for  studies of radio emission from clusters. 

Science data processing  will take place in an automated pipeline processor in real time. 
The on-line imaging uses neither natural nor uniform weighting, but instead uses an algorithm called preconditioning, which, together with multi-scale clean, gives a similar sensitivity to uniform weighting at small spatial scales, and a similar sensitivity to natural weighting at large spatial scales. So near-optimum sensitivity is obtained at all scales without needing to reweight the data. The $\sim10\, \mu$Jy/beam rms continuum sensitivity in 12 hours is approximately constant for beam sizes from 10 to 30 arcsec, then increases to $\sim20\, \mu$Jy/beam for a 90 arcsec beam and $\sim40\,  \mu$Jy/beam for a 3 arcmin beam.
 

One important key project of ASKAP will be EMU, the ``Evolutionary Map of the Universe'' (Norris et al 2011), 
an all-sky continuum survey.
The primary goal of EMU is to make a deep (10  $\mu$Jy/beam rms) radio continuum survey of the entire Southern Sky, extending as far North as $+30\deg$. EMU will cover roughly the same fraction (75\%) of the sky as the benchmark NVSS survey (Condon et al. 1998) 
, but will be 45 times more sensitive, and will have an angular resolution (10 arcsec) 4.5 times better. Because of the excellent short-spacing \emph{uv} coverage of ASKAP, EMU will also have higher sensitivity to extended structures such as cluster haloes.

\subsection{APERTIF: WODAN survey}

APERTIF, the new Phased Array Feed (PAF) system that will be installed on the Westerbork Synthesis Radio Telescope (WSRT), will dramatically increase,  at frequencies from 1.0 to 1.7 GHz,  the instantaneous field of view of the WSRT and its observing bandwidth (\eg Oosterloo et al 2010). Many beams can be formed simultaneously for each dish making it possible to image an area of about 8 square degree on the sky, which is an increase of about a factor 30 compared to the current WSRT. This entire field of view will be imaged with 15 arcseconds spatial resolution over a bandwidth of 300 MHz with a spectral resolution of about 4 km/s. The survey speed of APERTIF, and many of the other characteristics, will be very similar to ASKAP.

The extremely large field of view of APERTIF would enable the WODAN (Westerbork Observations of the Deep APERTIF Northern-Sky) project (\eg R\"ottgering et al. 2011). This project has been proposed with the aim to chart the entire accessible northern sky at 1400 MHz down to 10 $\mu$Jy rms and about 1000 deg$^2$ down to 5 $\mu$Jy. WODAN will be an important complement of the EMU project in the northern sky.

\section{Number of radio halos in the EMU+WODAN survey}

\begin{figure}
\includegraphics[width=0.5\textwidth]{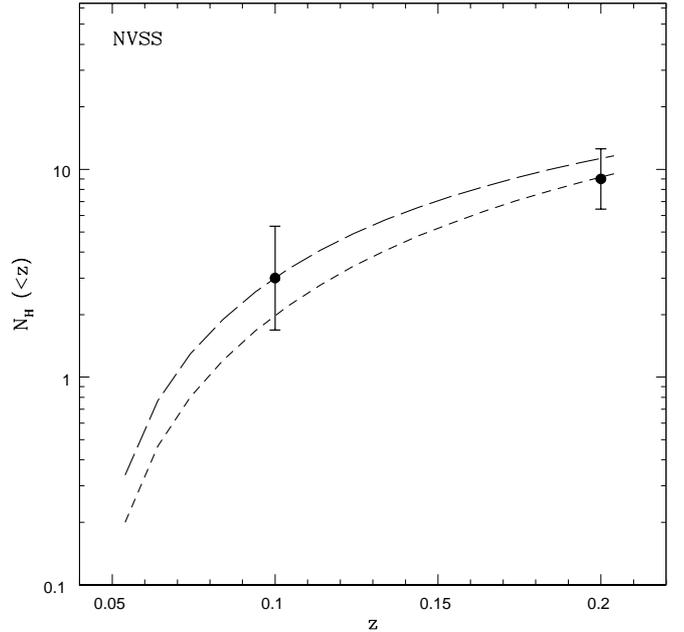}
\caption{
Integrated number of radio halos within a given redshift expected in the NVSS follow-up of the XBACS clusters (dashed line) compared
with the observed values (black points) within z=0.1 and z=0.2 taken from Giovannini et al. (1999).
The expected numbers of halos are computed by consider $f_{min}(z)$ given by Eq.~\ref{fmin} with $\xi_1=3$ (short dashed line), 
and by Eq.~\ref{fminhuub} with $\xi_2=10$ (long dashed line).}
\label{fig:NC_NVSS}
\end{figure}

\begin{figure*}
\begin{center}
\includegraphics[width=0.4\textwidth]{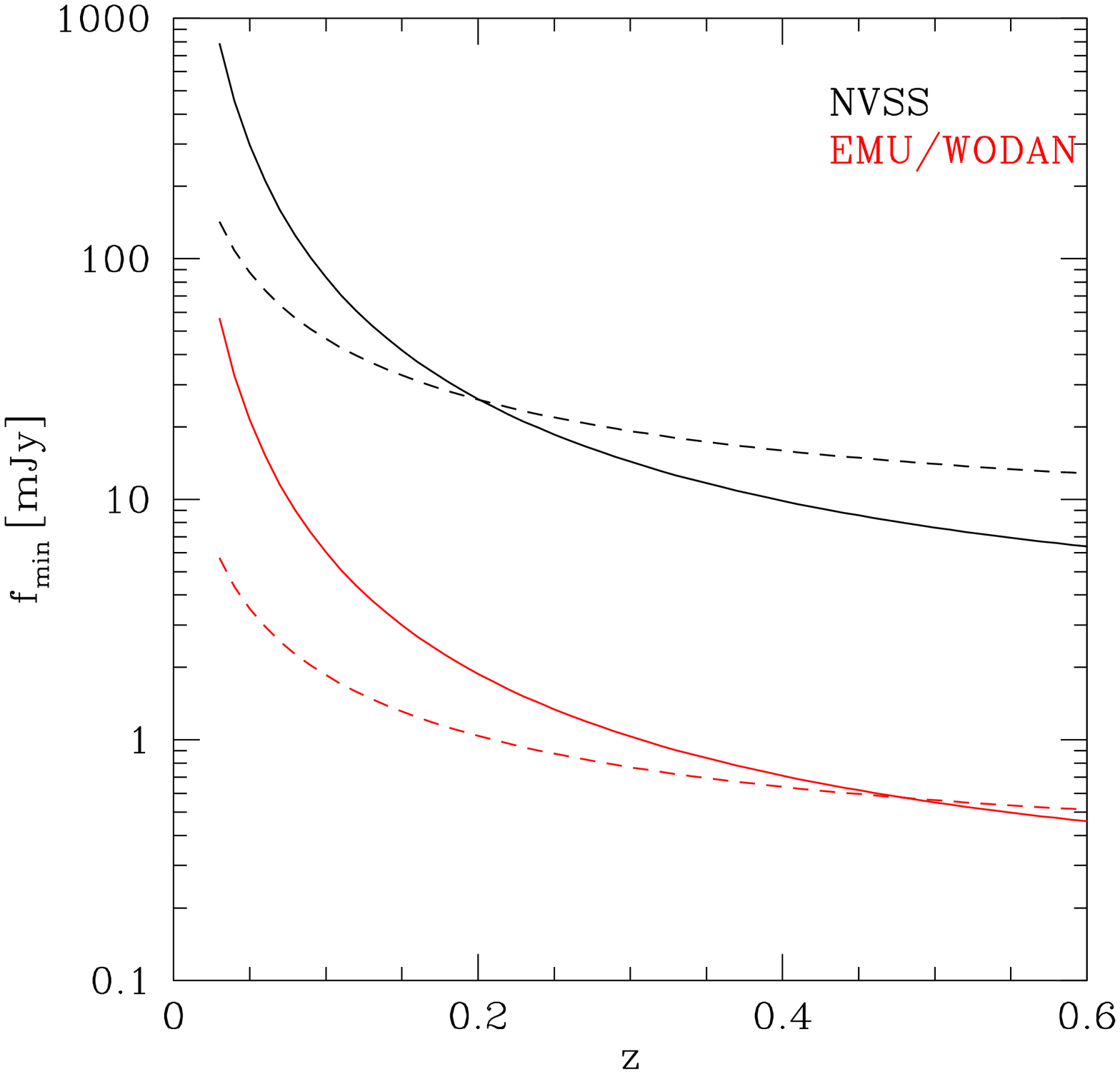}
\includegraphics[width=0.4\textwidth]{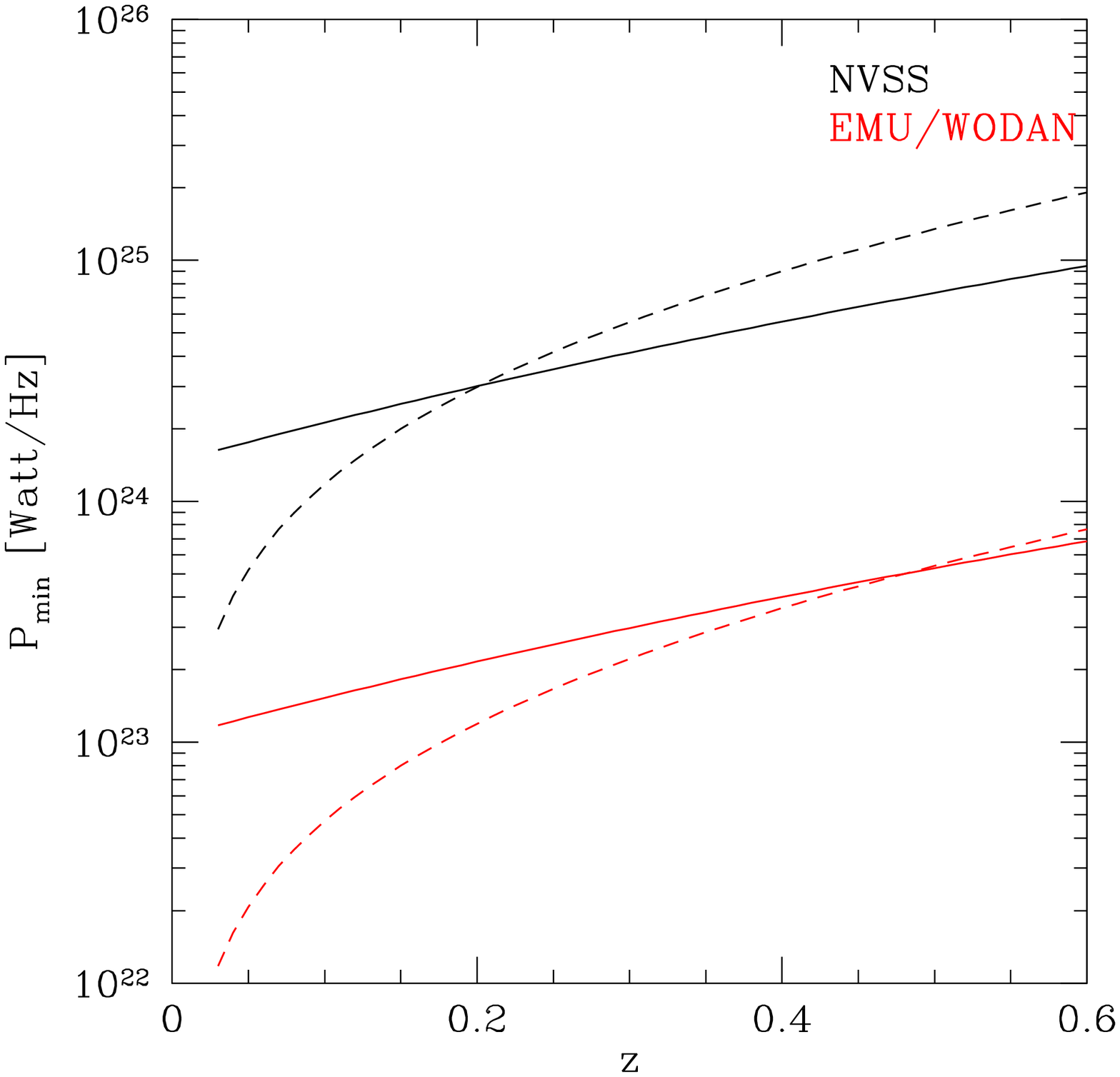}
\caption[]{
Minimum flux (left panel) and power (right panel) of detectable radio
halos at 1400 MHz obtained by Eq.~\ref{fmin} with $\xi_1=3$ (solid lines) and by Eq.\ref{fminhuub} with
$\xi_2=10$ (dashed lines). Calculations are shown for the NVSS (rms=$0.45$ mJy, $\theta_b=45$ arcsec; black upper lines)
and for EMU+WODAN (rms=$10\mu$Jy, $\theta_b=25$ arcsec; red bottom lines).
}
\label{Fig.flim_plim}
\end{center}
\end{figure*}

It has been shown that results based on calculations carried 
out according to Sect.2 are consistent with the observed increase 
of the fraction of clusters with radio halos with the cluster 
mass (or X-ray luminosity; Cassano et al.~2008a) and with the observed number 
of nearby radio halos (Cassano et al.~2006). 
In particular, Cassano et al. (2010b) showed that model expectations 
also produce the flux distribution of giant radio halos 
observed in the redshift ranges $0.044\leq z\leq 0.2$ 
(derived from the NVSS; Giovannini et al.~1999) 
and $0.2\leq z\leq 0.32$ (derived from the {\it GMRT radio halo survey}; 
Venturi et al.~2007, 2008).

The EMU+WODAN survey will explore the radio sky with a sensitivity $\sim$ 10 
times better than present surveys, making it possible to test models 
in a totally unexplored range of radio halo luminosities and masses of the host systems.
In this Section we shall derive the expected number of radio halos at 1400 MHz and 
explore the potential of the upcoming EMU+WODAN survey. 

\noindent
At this point it is important to estimate the minimum flux
of a radio halo (integrated over a scale of $\sim$ 1 Mpc) that 
can be detected in the survey. We will consider two possible approaches:
{\it i)} a brightness-based criterion and {\it ii)} a flux-based
criterion.

\noindent {\it i)} The criterion based on a threshold in brightness
guarantees that halos are detected in the images generated by
the survey. From this threshold we derive the flux (luminosity, z)
of radio halos that can be detected by assuming a spatial distribution of
their brightness. 
The brightness profile of giant radio halos is known to smoothly decrease with distance from the cluster 
center (\eg Murgia et al. 2009) implying that the outermost, low brightness, regions of halos are very difficult to detect. 
However what is important is the capability to detect at least, the brightest regions, of the radio halos.
Radio halos emit about half of their total radio flux within their half radius (Brunetti et al. 2007, Fig.~1).
Following Cassano et al. (2010b) we estimate the minimum flux of a halo, $f_{min}(z)$, that can be detected in the survey
by requiring that the mean halo brightness within half halo radius ($0.5~\theta_H$) is 
$\xi_1$ times the noise level in the map, \ie $0.5 f_{min}/N_{b}\approx \xi_1 F_{rms}$, where $N_b$ is the number 
of independent beams within $0.5\theta_H$ and $F_{rms}$ is the rms noise per beam.

\noindent This gives:

\begin{equation}
f_{min}(z)\simeq 1.2 \times10^{-4} \xi_1\,
\Big(\frac{\mathrm{F_{rms}}}{10 \mu\mathrm{Jy}}\Big)
\Big(\frac{100\,\mathrm{arcsec}^2}{\theta_b^2}\Big)
\Big(\frac{\theta_{H}^2(z)}{\mathrm{arcsec}^2}\Big)
\,\, [\mathrm{mJy}]\,,
\label{fmin}
\end{equation}

\noindent
where $\theta_{H}(z)$ is the angular size of radio halos at a given redshift in arcseconds and 
$\theta_b$ is the beam angular size of the survey in arcseconds. 

\noindent {\it ii)}
A second possible approach to derive $f_{min}$ is to assume that the halo is detectable
when the integrated flux within $0.5\,\theta_H$ gives a signal to noise ratio $\xi_2$. This is:

\begin{equation}
f_{min}(<0.5\,\theta_H)\simeq \xi_2\,\sqrt{N_b}\times F_{rms},
\label{fmin05}
\end{equation}

\noindent From Eq.~\ref{fmin05} it follows 

\begin{equation}
f_{min}(z)\simeq1.43\times10^{-3}\, \xi_2\,
\Big(\frac{\mathrm{F_{rms}}}{10 \mu\mathrm{Jy}}\Big)
\Big(\frac{10\,\mathrm{arcsec}}{\theta_b}\Big)
\Big(\frac{\theta_{H}(z)}{\mathrm{arcsec}}\Big)\, \, [\mathrm{mJy}]\,,
\label{fminhuub}
\end{equation}

In the following, for a better comparison with previous  works (Cassano et al. 2010b), we will consider {\it i)} as our reference approach,
and will give some results also based on {\it ii)}.

The number of radio halos with $flux\geq f_{min}(z)$
in the redshift interval, $\Delta z=z_2-z_1$, can be 
obtained by combining the RHLF ($dN_H(z)/dP(\nu_o)dV$) and $f_{min}(z)$:

\begin{equation}
N_{H}^{\Delta_z}=\int_{z=z_1}^{z=z_2}dz' ({{dV}\over{dz'}})
\int_{P_{min}(f_{min}^{*},z')}{{dN_H(P(\nu_o),z')}\over{dP(\nu_o)\,dV}}\,
dP(\nu_o)
\label{RHNC}
\end{equation}

\noindent
Estimating $\xi_1$ and $\xi_2$ is the more critical point in this procedure.
Considering case {\it i)}, in Cassano et al.~(2008a) we analysed VLA radio observations in D-array
configuration of empty fields where we injected fake radio halos
in the (u,v) plane of the interferometric data; the 
injected radio halos were placed at z=0.1 assuming a diameter $\sim$1 Mpc.
Cassano et al. concluded that radio halos become visible in the images
as soon as their flux approaches that in Eq.~\ref{fmin} with $\xi_1 \sim
1-2$. At redshift $z > 0.1$ the number of independent beams that cover
the half-radius region of the halos decreases, so based on our experience
a value $\xi_1 \sim 3$ yields a more reliable threshold if we are interested
in addressing the detection of halos over a wider redshift range,
\eg z=0-0.6.
Brunetti et al. (2007) and Venturi et al. (2008) injected ``fake'' radio halos in the (u,v) 
plane of GMRT data to evaluate the flux of detectable radio halos. Based on their
findings (Fig.~3 in Brunetti et al. 2007) we adopt Eq.~\ref{fminhuub} with $\xi_2\simeq10$.


\noindent
As a consistency check we derive the expected number of radio halos to be detected in the NVSS 
up to $z\leq0.2$ and compare them with those derived from the cross-correlation of the XBACs sample (Ebeling et al. 1996)
with the NVSS (from Giovannini et al. 1999).
This is shown in Fig.~\ref{fig:NC_NVSS}: values of $\xi_1\sim 3$ and $\xi_2\sim 10$
gives a predicted number of radio halos consistent with that observed in the NVSS
\footnote{We note that case {\it i)} gives a more fair comparison because Giovannini et al. (1999)
selected halos and candidate halos from the inspection of the NVSS images.}.
In the following we will use Eq.~\ref{fmin} with $\xi_1=3$ to compute the minimum flux 
of detectable radio halos, and we will provide also expectations by considering 
Eq.~\ref{fminhuub} with $\xi_2=10$ in a number of cases. 

Figure.~\ref{Fig.flim_plim} shows $f_{min}$ of giant radio halos as a function of redshift (left panel), 
and the corresponding minimum radio luminosity (right panel), obtained according to case {\it i)} and {\it ii)}.
Calculations are shown for the NVSS (black upper lines) and for the EMU+WODAN survey (red lower lines). 
For the EMU+WODAN survey we assume $rms=10\,\mu$Jy and $\theta_b=25$ arcsec \footnote{We note that such noise level is of the same order of the
confusion noise expected in this configuration, however it can be reached after subtraction of uv components of the sources detected at more than 100 $\sigma$ in the survey images obtained at the highest resolution.}.

\begin{figure*}
\begin{center}
\includegraphics[width=0.4\textwidth]{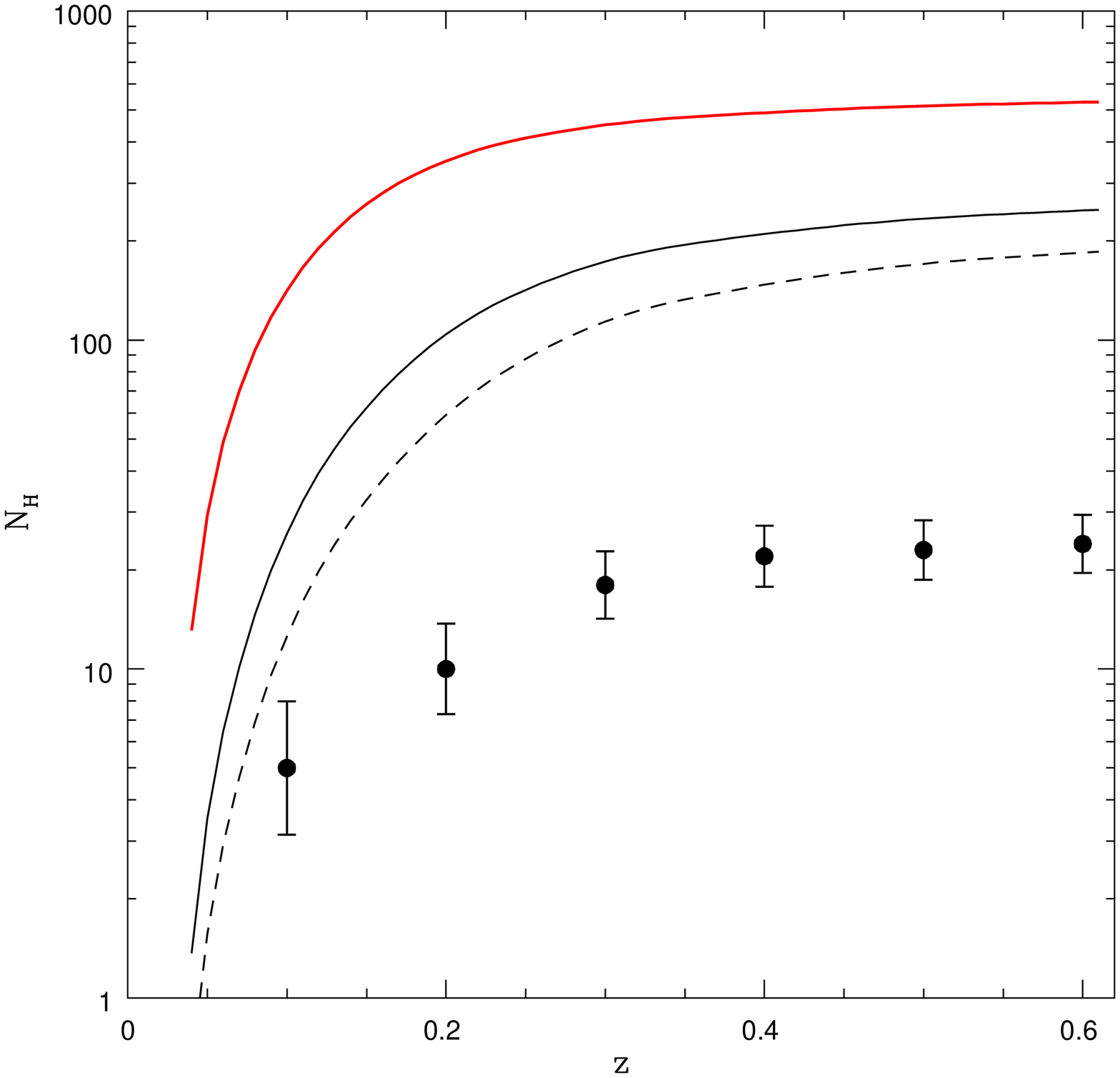}
\includegraphics[width=0.4\textwidth]{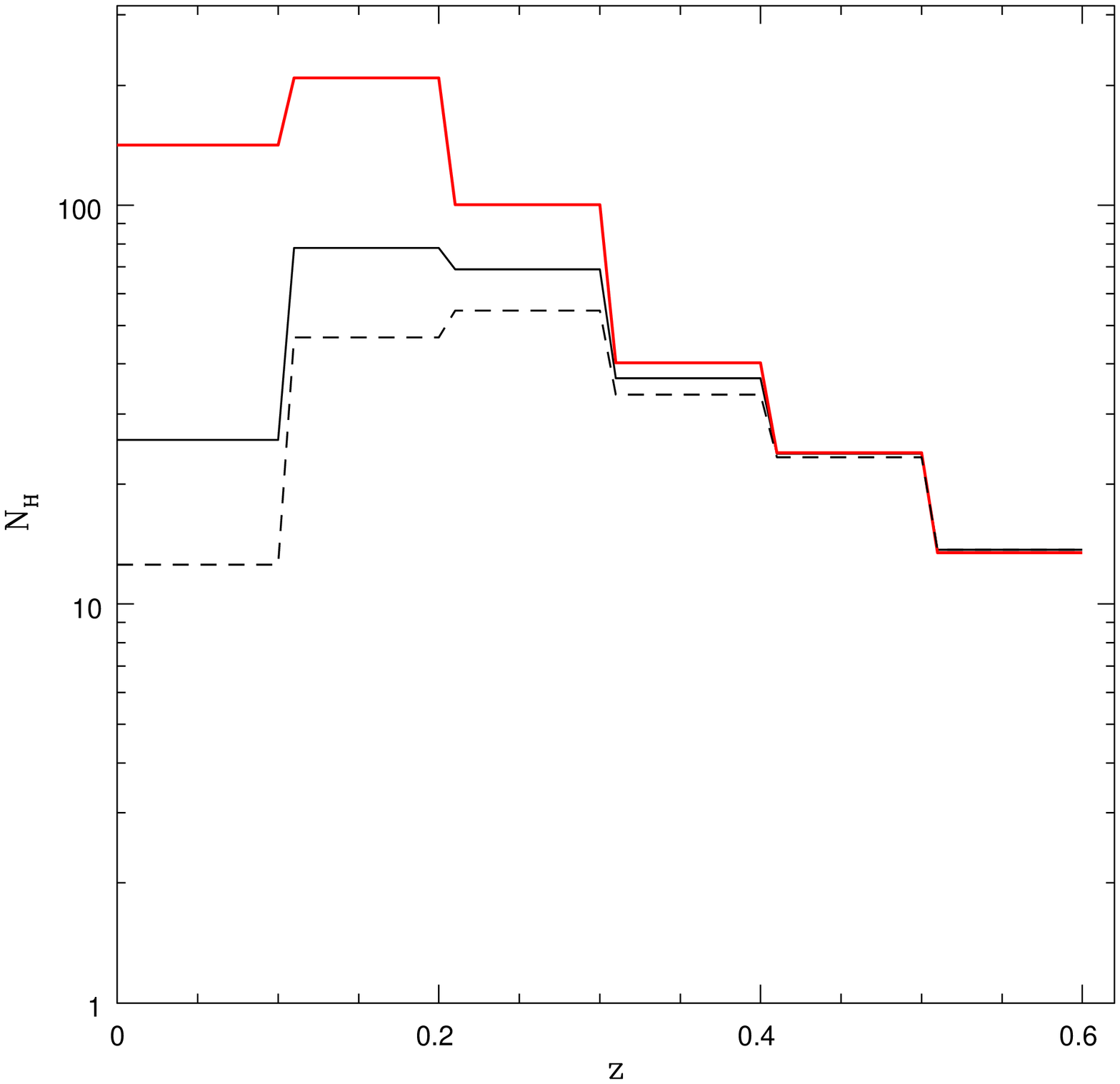}
\caption[]{{\it Left Panel}: expected integral number (all-sky) of radio halos as a function of redshift in EMU+WODAN. {\it Right Panel}: expected number of RH (all-sky) in redshift intervals in EMU+WODAN. The reference set of model parameters ($b=1.5$, $B_{<M>}=1.9\,\mu$G, $<M>=1.6\times 10^{15} M_{\odot}$ and $\eta_t=0.2$) is assumed.
In both panels black lines show the total number of halos (``turbulent''+``off-state'' halos, solid) and of ``turbulent'' halos (dashed) obtained by considering Eq.\ref{fmin} with $\xi_1=3$ to derive $f_{min}(z)$; while red upper lines show the total number of halos (``turbulent''+``off-state'' halos) obtained by adopting Eq.\ref{fminhuub} with $\xi_2=10$.
Points show the integral number of radio halos observed so far.}
\label{Fig.EMU_RHNC}
\end{center}
\end{figure*}

\noindent 
In Fig.~\ref{Fig.EMU_RHNC}, ({\it left panel}) we show the all-sky number of radio halos 
expected in the EMU+WODAN survey.
We consider both giant radio halos that originate from turbulent re-acceleration in merging
clusters and ``off-state'' halos assuming the optimistic case b) in Sect.~3 (see caption for details). 
We consider the flux limit derived according to Eq.~\ref{fmin} with $\xi_1=3$ (black lines)
and report also the total number of halos expected according to Eq.~\ref{fminhuub} with $\xi_2=10$ 
(red lines).

\noindent
We note that ``off-state'' halos are expected to contribute significantly
(about 30\%) to the total number of radio halos that are expected in
the EMU+WODAN survey. Their contribution is larger at lower redshift ($z<0.3$).


\noindent
We predict that the EMU+WODAN survey will detect up to 100-200 radio halos in 
the redshift range 0--0.6. This will increase the number of presently known giant radio halos 
by about one order of magnitude. About 2/3 of these radio halos are expected 
in the redshift range $0.1-0.4$ (Fig.\ref{Fig.EMU_RHNC}, {\it right panel}). 
The number of radio halos expected in the EMU+WODAN surveys increases by a factor $\sim2$ 
when considering the flux limit given by Eq.~\ref{fminhuub}.
This increase is due to an increasing number of ``off-state" radio halos that become detectable 
in low redshift ($z<0.2$) clusters according to this prescription.

\begin{figure}
\begin{center}
\includegraphics[width=0.5\textwidth]{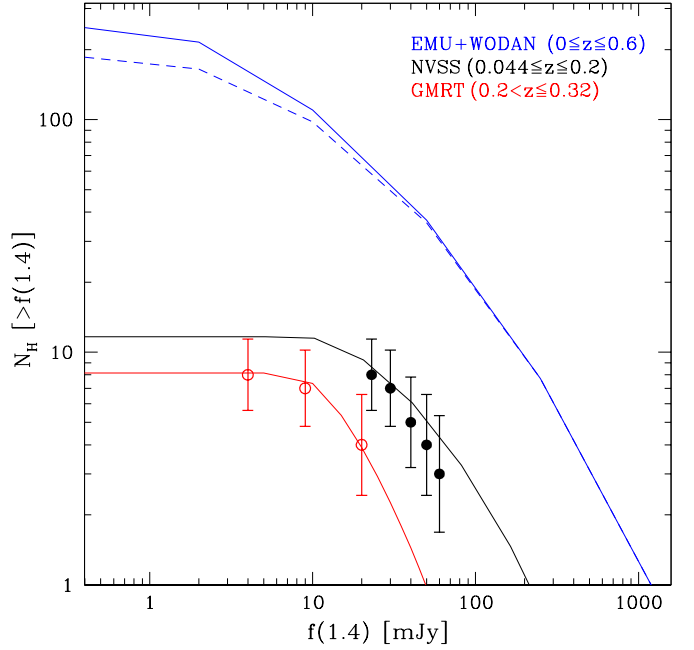}
\caption[]{All-sky number distribution of radio halos within $z<0.6$ as a function of the radio flux at $1.4$ GHz as expected in the EMU+WODAN surveys
(blue lines). 
For comparison, in the same figure, the expected distribution of radio halos in the redshift range $0.044<z<0.2$, compared with that observed in the NVSS (black lines and symbols) and in the redshift range $0.2<z<0.32$ compared with that observed in the ``GMRT RH Survey'' (red lines and symbols) are also reported. The blue dashed line shows the distribution of turbulent generated halos.}
\label{Fig.EMU_RHNC_flux}
\end{center}
\end{figure}

Finally, in Fig.\ref{Fig.EMU_RHNC_flux} we report the all-sky 
number distribution of the radio halos detectable by EMU+WODAN 
as a function of the radio flux.
``Off-state'' halos 
contribute potentially at smaller radio fluxes, $f_{1.4}<10$ mJy, 
\ie at fluxes presently accessible only to deep pointed observations. 
To further highlight the potential improvement that will be provided 
by EMU+WODAN with respect to the present statistics 
of radio halos, in the same figure we report the number of radio halos 
detected in the NVSS survey (by inspection of XBAC clusters up 
to redshift $0.2$, Giovannini et al 1999) 
and in the GMRT RH Survey (Venturi et al. 2007, 2008; in the redshift 
range 0.2-0.32). 
For a sanity/consistency check, the number of radio halos in present surveys 
are also compared with model expectations derived for turbulent giant
radio halos according to Eqs.\ref{fmin}--\ref{RHNC} and by
taking into account the specification of these surveys.

\section{Comparison between LOFAR and EMU surveys}

The revolutionary radio telescope LOFAR will carry out surveys 
between 15 MHz and 210 MHz with unprecedented sensitivity and 
spatial resolution (\eg R\"ottgering et al. 2006), 
providing a breakthrough in the exploration of the Universe
at the low radio frequencies.
In particular the {\it Tier 1} ``Large Area'' survey of the
northern sky is planned to reach sensitivities $\sim$0.1 mJy/beam
in the frequency range 120-190 MHz. The combination of observing
frequency and sensitivity to diffuse emission 
of {\it Tier 1} make this survey the most sensitive survey in the
pre-SKA era 
for the exploration of non-thermal radio emission
from galaxy clusters and large scale structure.

\noindent
Based on the hypothesis that giant radio halos originate
from turbulent re-acceleration in merging clusters, Cassano et
al.~(2010b) predict the discovery of about 400 giant radio halos 
at redshifts $\leq0.6$ from the analysis of the {\it Tier 1} survey data.
This would increase the statistics of these sources by a factor 
of $\sim 20$ with respect to present day surveys. 
Remarkably about 1/2 of these halos are expected with a synchrotron spectral 
index between 250-600 MHz of $\alpha > 1.9$, and consequently they  
would brighten only at low frequencies, unaccessible to both 
present observations and future observations with ASKAP and APERTIF.

The surveys planned with the ASKAP array and with the APERTIF onboard of
WSRT will provide information complementary to that coming from
LOFAR. These surveys are planned to reach a sensitivity similar to that 
of the {\it Tier 1} survey in the case of extended emission with radio spectral
index in the range $\alpha\sim 1-1.2$.
Their synergy with LOFAR will add considerable value 
to discriminate between different physical origin of
giant radio halos in galaxy clusters and, in general, to constrain
the evolution of
cluster non-thermal components.

In this Section we shall focus on the comparison between expectations 
for the statistics of radio halos in LOFAR and EMU+WODAN. 
LOFAR surveys and EMU will overlap by about one steradiant, 
while WODAN will be looking at the same sky as LOFAR.


\noindent
Here we calculate the expected number of giant radio halos in the LOFAR 
{\it Tier 1}
survey following Cassano et al.~(2010b).
We upgrade these calculations (i) by using {\it merger trees}
from Monte Carlo simulations of cluster formation history
with increased statistics with respect to what 
previously done (\ie increased number of trials), and (ii) by adopting a 
tighter grid to sample the values of the steepening frequency of
radio halos, $\nu_s$.
In addition, following Sect.3 we include the contribution to the halos statistics at 
120-190 MHz from ``off-state'' halos in more relaxed clusters. 
We adopt a spectral index $\alpha = 1$ to estimate the radio luminosity 
of these halos at lower frequencies.

\begin{figure*}
\begin{center}
\includegraphics[width=0.4\textwidth]{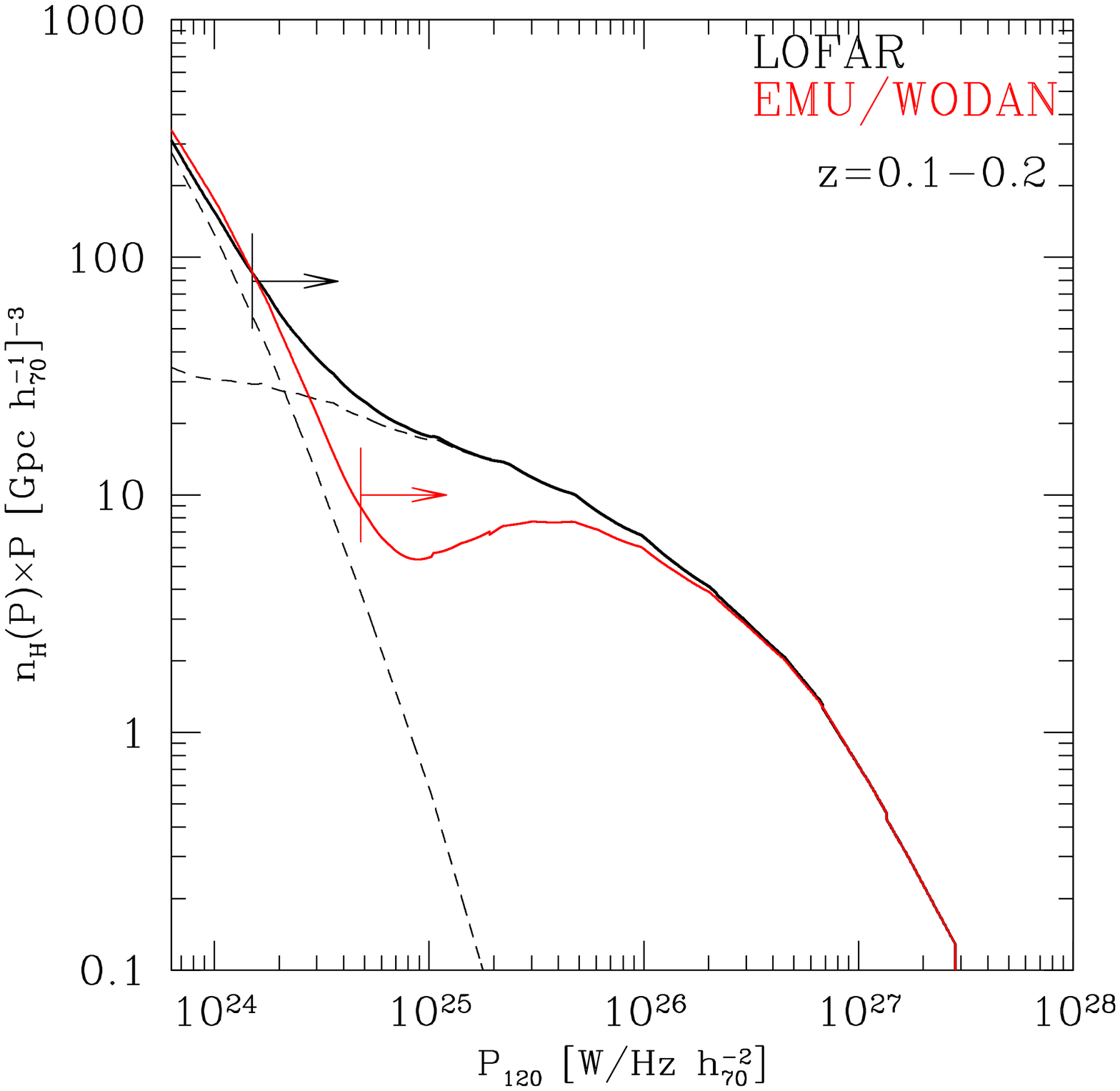}
\includegraphics[width=0.4\textwidth]{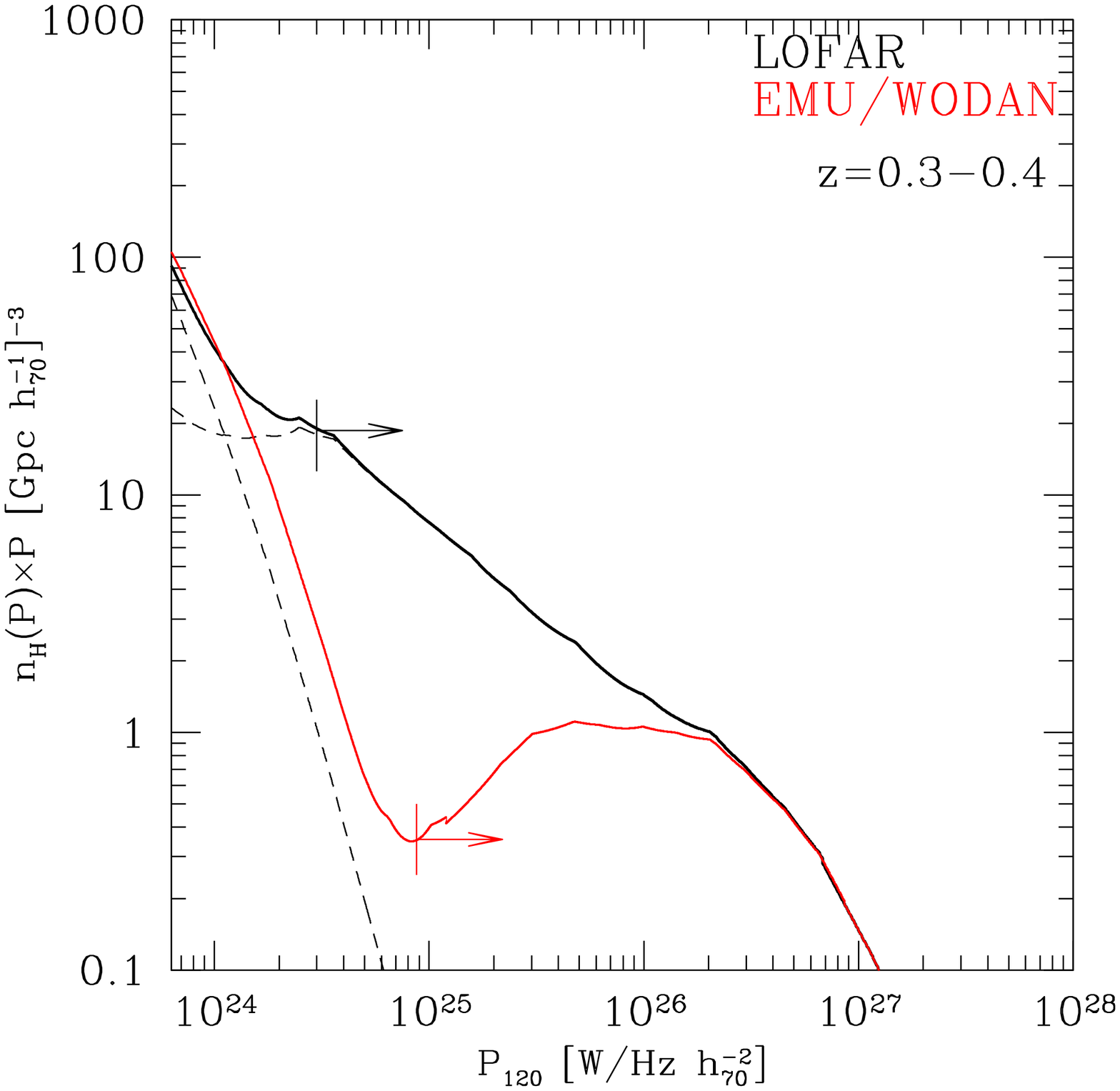}
\caption[]{RHLF of halos with $\nu_s\ge120$ MHz (black solid lines) and that of halos with $\nu_s>1000$ MHz (red lines) in two different redshift ranges z=$0.1-0.2$ (left panel) and z=$0.3-0.4$ (right panel). As an example, dashed lines (in both panels) show the individual contributions of ``turbulent'' radio halos and ``off-state'' halos to the LOFAR RHLF. The black and red arrows show the LOFAR and EMU+WODAN sensitivities, respectively, at the considered redshifts.}
\label{Fig.LOFAR_EMU_RHLF}
\end{center}
\end{figure*}

LOFAR and EMU+WODAN will observe 
different populations of giant radio halos, because LOFAR will explore also the population
of halos with steepening frequency $\nu_s \leq 1$ GHz.
As a first step we calculate the RHLF of these two population
of halos.
As a relevant example, Fig.~\ref{Fig.LOFAR_EMU_RHLF} shows the RHLF of halos 
with $\nu_s>120$ MHz (black solid lines) and that of halos 
with $\nu_s>1$ GHz (red lines) in two different 
redshift ranges z=$0.1-0.2$ (left panel) and z=$0.3-0.4$ (right panel). 
For a prompt comparison between the two populations, the luminosities of halos with $\nu_s>1$ GHz, 
which emit in the frequency band spanned by EMU+WODAN, are extrapolated at 120 MHz by assuming 
a spectral index $\alpha\approx 1.3$. 
The difference between the two RHLF is maximized at higher redshift 
where the RHLF of high-frequency halos shows a dip at radio luminosities
$\approx 1\times 10^{25}$ Watt/Hz (at 120 MHz). It reflects the
difficulty of generating high-frequency giant halos via turbulent re-acceleration
in the presence of stronger inverse Compton losses (higher z).
Indeed those high-frequency halos can be generated at higher redshift only
by very massive merging events, which however preferentially
produce halos with larger radio luminosities (\eg Eqs.~4--5).
On the other hand, we find that radio halos emitting at lower frequencies 
can still be generated efficiently at higher redshifts.
Consequently the detection at low frequencies of a number of radio halos 
with luminosities $\sim 10^{25}-10^{26}$W/Hz 
in excess of that of radio halos observed at higher frequencies
with luminosities $\sim 5\times 10^{23}-5\times 10^{24}$W/Hz would
confirm our theoretical expectations; 
the sensitivities of the LOFAR {\it Tier 1} 
survey and of the planned EMU survey
will be suitable to perform this test (Fig.~\ref{Fig.LOFAR_EMU_RHLF}).

\noindent
At lower luminosities both the high-frequency and low-frequency
RHLF are dominated by the contribution of ``off-state'' halos.
We note that the number density of these halos at lower frequencies
is slightly smaller than that at higher frequencies (red curves are slightly higher
than the black curves, Fig.~\ref{Fig.LOFAR_EMU_RHLF}) because
at lower frequencies an increasing number of clusters can generate 
giant radio halos via turbulent re-acceleration. In other words,
some of the clusters hosting low luminosity ``off-state'' (``secondary'') halos become ``on-state''
(turbulent) when observed at LOFAR frequencies, and migrate from low to high radio luminosities
in the LOFAR luminosity function.
Fig.~\ref{Fig.LOFAR_EMU_RHLF} shows that both LOFAR and EMU+WODAN
will be able to detect ``off-state'' halos (assuming the optimistic case b) in Sect.3).

\begin{figure*}
\begin{center}
\includegraphics[width=0.4\textwidth]{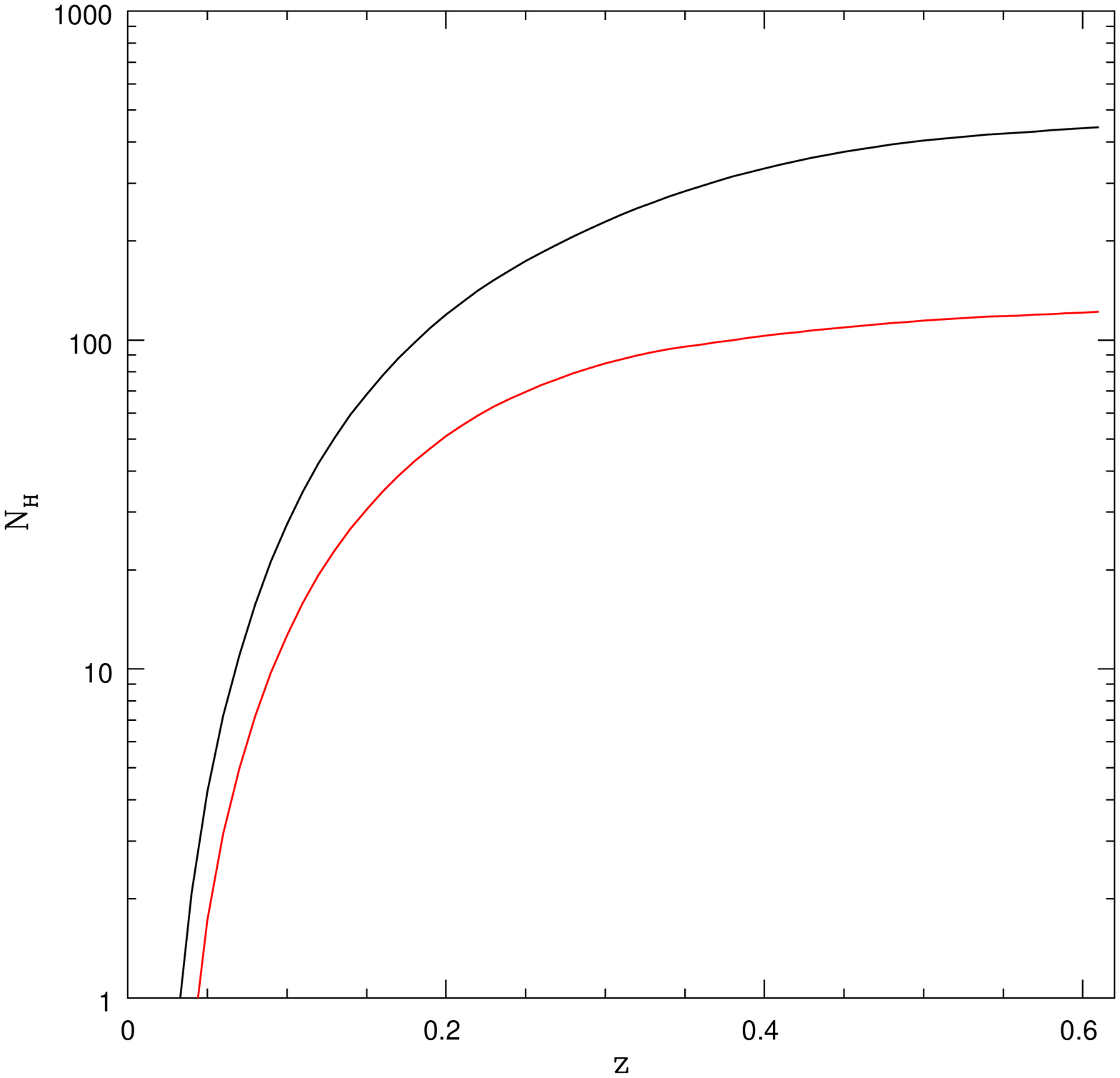}
\includegraphics[width=0.4\textwidth]{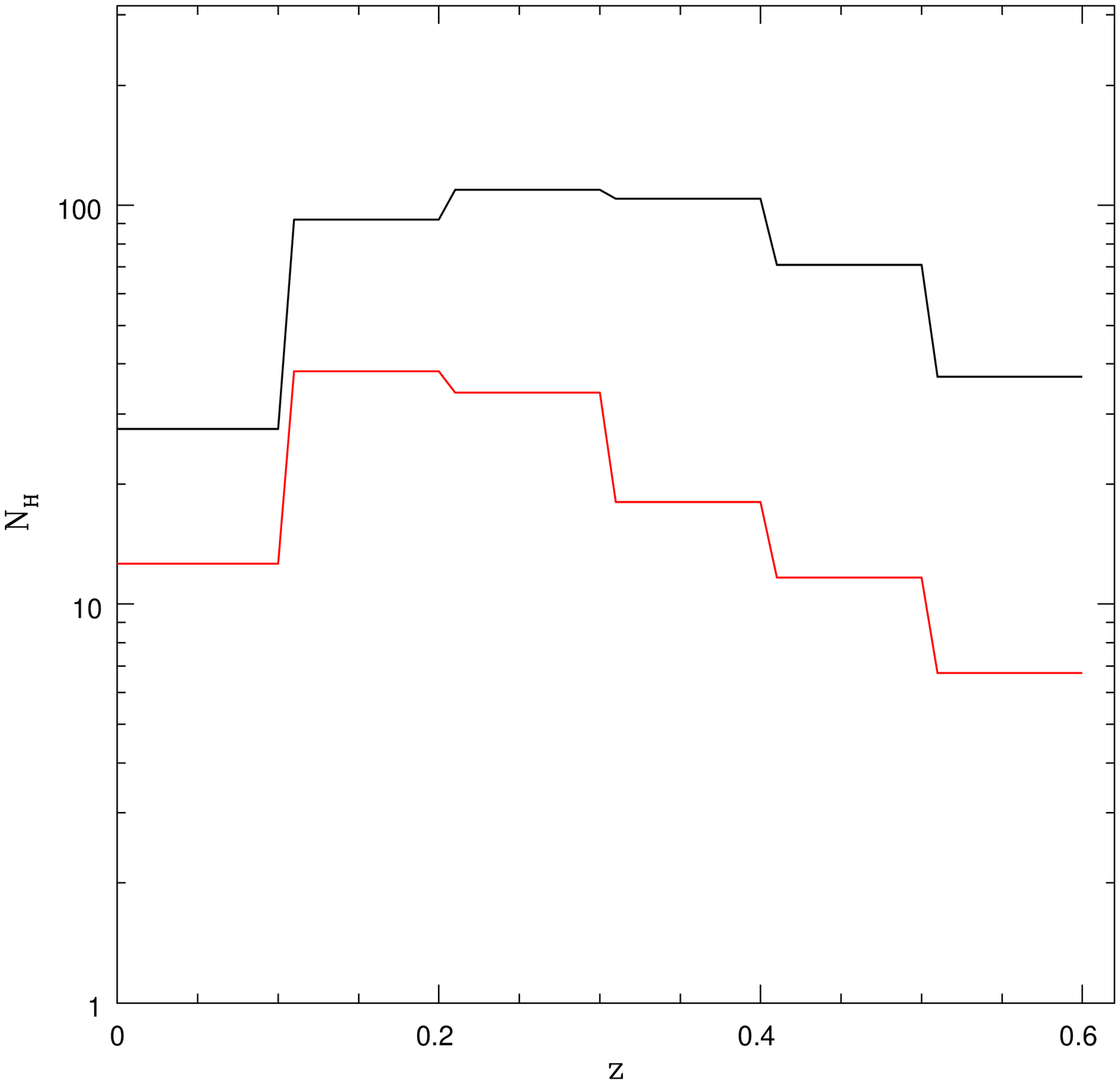}
\caption[]{{\it Left Panel}: integral number of radio halos as a function of redshift in the WODAN (red line) and LOFAR (black line) surveys. {\it Right Panel}): RH distribution in redshift intervals in the WODAN (red line) and LOFAR (black line) surveys.}
\label{Fig.LOFAR_EMU_RHNC}
\end{center}
\end{figure*}

Since LOFAR and WODAN will be looking at the same sky with the same 
sky coverage, here we report a comparison between the expected number of radio halos 
in these two surveys. To derive the number of radio halos 
we follow the procedure in Sect.~5, assuming a threshold value
$\xi_1 \sim 3$ for both LOFAR and WODAN.
In Fig.~\ref{Fig.LOFAR_EMU_RHNC} ({\it left panel}) we show 
the integral number of giant radio halos (in the LOFAR sky, \ie in $\sim 20214$ square degrees)
as a  function of redshift in the WODAN (red line) and LOFAR (black line) 
surveys, and in Fig.~\ref{Fig.LOFAR_EMU_RHNC} ({\it right panel})
we report the number of radio halos in redshift intervals. 
The LOFAR {\it Tier 1} survey is expected to detect about 500 radio halos, $\sim 4$ 
times more than WODAN.
This is because of the better sensitivity of the LOFAR
survey (see also Fig.~\ref{Fig.LOFAR_EMU_RHLF}) and from the possibility
to detect low-frequency radio halos (\ie $\nu_s <$ GHz) at the frequencies
spanned by LOFAR.
The difference between the two surveys increases with redshift mainly due 
to the increasing number of low-frequency radio halos that are generated
in cluster mergers at higher redshifts.

\noindent
Giant radio halos with very steep spectrum (low-frequency halos) are
only expected in the framework of the {\it turbulent re-acceleration} 
models (\eg Cassano et al. 2006; Brunetti et al. 2008) thus unveiling
a substantial population of these radio sources 
will promptly discriminate among different models proposed for the origin of diffuse radio
emission in galaxy clusters.

\section{Discussion and conclusions}

In this paper we have presented results from Monte Carlo simulations to model 
the formation and evolution of giant radio halos in the framework 
of the merger-induced particle acceleration scenario (see Sec.~2) 
and extended previous calculations by including the contribution 
of secondary electrons generating ``off-state'' halos 
in more relaxed galaxy clusters (Sect.~3).
To combine these two mechanisms we follow 
a phenomenological approach in which we assume that those clusters 
where turbulence is not sufficient to generate radio halos
emitting at the observing frequency, $\nu_o$, host
``off-state'' halos generated only by the emission from secondary electrons. 
We assume that presently observed giant radio halos
are mainly driven by turbulent re-acceleration in merging clusters and
constrain the level of the ``off-state'' halos using limits derived for ``radio quiet'' galaxy clusters (Brunetti et al. 2007; 
Brown et al. 2011).
Under this assumption ``off-state'' halos are faint with luminosities typically
$\sim 10$ times smaller than those of giant radio halos in turbulent/merging
clusters. On the other hand, these ``off-state'' halos are expected to be
more numerous than turbulent halos and thus they may contribute significantly to the number of radio halos
in future radio surveys.

The most important expectation of turbulent re-acceleration
scenarios is that the synchrotron spectrum of radio halos should become 
gradually steeper above a frequency, 
$\nu_s$, that is determined by the competition between acceleration and
energy losses and that is connected to the energetics of the merger events 
that generate the halos (\eg Fujita et al.~2003; Cassano \& Brunetti 2005).
Consequently, in this scenario the population of radio halos is expected to be made of
a complex mixture of sources with different spectra, with massive (and hot) 
clusters that have a tendency to generate halos with spectra flatter than those 
in less massive systems. Contrary to turbulent halos, ``off-state'' halos are expected with
power-law spectra with fairly similar slopes. Consequently,
 surveying the sky at different radio frequencies and with appropriate
sensitivities allows to disentangle these two populations. 

In Sects. 2-3 we derive the expected radio halo luminosity 
functions (RHLF) at frequency $\nu_o$, that account for radio halos originated 
from turbulent re-acceleration, with steepening frequency $\nu_s \geq \nu_o$,
 and ``off-state'' radio halos.
As a relevant case we discuss the RHLF at $\nu_o=1.4$ GHz.
The RHLF is characterized by a flattening at low radio luminosity,
$P \sim {\rm few} \times 10^{23}-10^{24}$W/Hz, 
due to the decrease of the efficiency of turbulent acceleration 
in less massive systems, and by an upturn at lower radio luminosities
due to the contribution of ``off-state'' halos.
The flattening/dip is expected to become more prominent
at higher redshift due to the increase of IC losses that quench the
acceleration process especially in lower massive systems.

Future radio surveys have the potential to constrain the formation
and evolution of halos with cosmic time allowing for detailed tests of models.
Specifically, in Sect. 5 we derive the expected number of radio halos in the EMU+WODAN
survey.
The EMU+WODAN surveys will probe the radio sky in the frequency range
1-2 GHz with a sensitivity 10 times 
better than present surveys. They will allow comparison of model predictions and observations in a totally unexplored range of radio halos 
luminosities and masses of the hosting clusters. A critical point in our paper is to derive a meaningful estimate
of the sensitivity of these surveys to radio halo emission at different redshifts.
We explore two possible ways, one based on a brightness threshold of the radio emission
and another one based on a threshold in flux density. Threshold values in both cases have been
estimated from the injection of ``fake'' radio halos in existing surveys (the GMRT RH Survey and the
NVSS) under the assumption that the brightness distribution of presently known halos is representative. 
A better determination of the sensitivity of EMU and WODAN will become available only
when ASKAP and APERTIF on WSRT will start the commissioning phase.
Despite the uncertainties on survey sensitivities, the expected number  
of radio halos highlights the potential of the future EMU+WODAN surveys. 
By using a brightness based criterion for the detection of halos and 
assuming the expected sensitivity of EMU (Norris et al. 2011) and WODAN
(R\"ottgering et al. 2011), rms $\sim 10\, \mu$Jy/b, we predict that these surveys will potentially
discover up to 200 new giant radio halos at redshift $z\le0.6$.
Most of these halos are predicted in the redshift
range $z \sim 0.2-0.4$.
This will increase the present number of known radio halos 
by almost a factor 10. The number of halos expected in these surveys
further increases if a flux based threshold is used to estimate the sensitivity
of the surveys. In particular, according to this method, more radio halos
can be discovered at $z<0.2$ with respect to the case of a threshold based on brightness. 

\noindent The leap forward of these surveys with respect to present observations
is due to the fact that the increasing sensitivity will make possible the detection of both
turbulent radio halos associated with less massive systems (or with $L_X\ltsim 2\times 10^{44}$ erg/s) 
and of ``off-state'' halos in more relaxed clusters.
We derive also the flux distribution of expected radio halos, 
showing that ``off-state'' halos contribute 
at fluxes $f_{1.4}<10$ mJy that are presently accessible 
only to deep pointed observations; still no clear detection of these halos has been obtained so far.

The most important step in our understanding of the physics of radio halos
is expected from surveys at lower frequencies and from their combination with surveys at higher
frequencies. In Sect.6 we compare model expectations for the statistics of
giant radio halos at lower and higher frequencies.
We upgraded previous calculations used to derive the number of radio halos at lower frequencies, $\nu_o \sim
120$ MHz, (from Cassano et al. 2010b)
{\it i)} by using Monte Carlo with improved statistics, and {\it ii)} by adopting
a scenario based on two-populations of halos (as in Sect. 5, ``off-state'' and ``turbulent'' halos).
These calculations allow us to explore the potential of the
synergy between surveys at lower and higher frequencies with LOFAR and EMU/WODAN, respectively. 
According to the model based on turbulent re-acceleration in galaxy 
clusters different population of giant radio halos should become visible
at different frequencies.
Based on our hypothesis of two-populations of halos, also the
population of ``off-state'' halos changes with frequencies 
because more clusters generate radio halos via turbulent re-acceleration
at lower frequencies.
We find that the LOFAR {\it Tier 1} survey will detect about 4 times
more halos than the WODAN survey, thanks to its better sensitivity and
to its lower observing frequency that allows the detection of turbulent radio halos
that become more frequent and luminous at lower frequencies. The majority of radio halos with very steep spectrum (\ie lower values of $\nu_s$)
in the LOFAR {\it Tier 1} survey are predicted with luminosities $P_{120} \sim 10^{25}-10^{26}$W/Hz
at 120 MHz and do not have counterparts detectable at higher frequencies.
As a consequence, the RHLFs at 1.4 GHz exhibit a dip/flattening (at the luminosities
$P_{1.4} \sim 10^{25}-10^{26} \times (1400/120)^{-1.3}\,$ W/Hz $\sim  
{\rm few} \times 10^{23}-10^{24}$W/Hz at 1.4 GHz). The comparison
between RHLFs at different frequencies is thus very important as it allows to
promptly unveil the existence of different populations of radio halos. For instance,
we have shown that comparing for radio halos in the LOFAR and in the WODAN surveys 
will allow a prompt test of the existence of radio halos with extremely steep spectra.

\noindent
We predict that both the LOFAR and EMU+WODAN surveys will have the potential
to unveil ``off-state'' radio halos, opening the possibility to study
diffuse Mpc-scale emission also in more relaxed systems.
This is currently impossible with present radio surveys, and will allow to better understand the effect of cluster mergers
on the evolution of non-thermal radio emission in galaxy clusters.

\noindent More generally, LOFAR and EMU+WODAN surveys will allow to readily discriminate between
a turbulent re-acceleration and a purely ``hadronic'' origin of radio halos. Indeed, if we assume that turbulence does not
play a role, and all halos are of ``hadronic'' origin, the luminosity function of ``hadronic'' halos in  Fig.\ref{Fig.LOFAR_EMU_RHLF} should
be boosted up by at least one order of magnitude to explain present number counts, implying a very larger number of halos at lower luminosities. 
Although a quantitative comparison between predictions derived by different authours is difficult because of the very different hypothesis
on (poorly constrained) model parameters\footnote{See, for example, the large discrepancies between  Sutter \& Ricker (2011) and Zandanel et al. (2012) which attempt to derive expectations using the ``hadronic'' model with different assumptions on the B-mass scalings and cosmic ray dynamics.}, we stress that the most important difference between the two scenarios is in the spectral shape of halos and in its consequences. Indeed, re-acceleration models lead to the {\it unique} predictions of complex spectra, whereas in the ``hadronic'' models the spectra of secondary particles (and thus the synchrotron spectra) are rather power-laws which extend in principle to very high energies (frequencies). The important consequences are that in re-acceleration models {\it i)} a still hidden population of radio halos with very steep spectra is predicted to glow up at low radio frequencies (\eg Cassano et al. 2006; Brunetti et al. 2008; Cassano et al. 2010b) and {\it ii)} the shape of the luminosity function of giant radio halos depends on the observing frequency (Fig.\ref{Fig.LOFAR_EMU_RHLF}); both predictions will be readily tested by future surveys at different frequencies.

\subsection{Model Simplifications}
Our calculations provide a first step for interpreting future surveys at the light of models that combine
turbulent re-acceleration of relativistic particles and the generation of secondary electrons in the ICM.
The most important simplification in our approach is that the evolution of magnetic field in the ICM does not
account for the possible connection with the level of turbulence, yet the magnetic field is simply anchored
to cluster mass at any redshift. Present data do not show evidence for a direct connection between the magnetic field
and the cluster dynamics and turbulence (Govoni et al. 2010; Bonafede et al. 2011) thus for the aim of the present
paper we prefer to keep the models as simple as possible. In this paper, we have shown results based on a reference set of
model parameters ($<B>$,b and $\eta_t$). According to Cassano et al. (2006) we expect that the general results given in the present
paper should be poorly dependent on the adopted parameter values. The expected number of radio halos in a given survey is expected
to change by a factor of about 2 when considering different values of parameters (see also Figs.14 and 15 in Cassano et al. 2006).
Our two-population scenario for halos does not consider self-consistently the evolution of protons and of their secondaries in the ICM, rather we use
a phenomenological approach based on the two ``extreme situations'' where halos originate from turbulence and pure hadronic
collisions, respectively. Brunetti \& Lazarian (2011) modeled the acceleration of relativistic protons and their secondaries in compressible
MHD turbulence in a self-consistent way showing that radio halos become bright when turbulence is generated in the ICM and gradually
evolve into fainter ``off-state'' halos when turbulence is dissipated. This complex evolution occurs in a time-scale shorter than the life-time
of halos and of the hosting clusters, thus our approach still provides a valuable way to model the basic statistical properties of halos (see discussion in Cassano et al. 2010b). As a final remark, in this paper we focussed only on the case of Mpc-sized radio halos. In reality, smaller halos could be generated in dynamically disturbed and relaxed systems by several mechanisms, including turbulent re-acceleration in sloshing cores and in the region of AGN-driven bubbles (Cassano et al. 2008b; Mazzotta \& Giacintucci 2008; ZuHone et al. 2012), reconnection regions (Lazarian \& Brunetti 2011) and hadronic collisions (Pfrommer \& En\ss lin 2004; Keshet \& Loeb 2010). All these halos constitute an additional population of diffuse synchrotron source in galaxy cluster to investigate with future radio surveys.

\noindent Finally, it's worth mentioning that there are a few radio halos found in clusters with X-ray luminosity lower than that of typical radio-halo clusters, that are over-luminous in radio (by about one order of magnitude) with respect to the radio-X-ray luminosity correlation (\eg Giovannini et al. 2011 and ref. therein). These halos have radio luminosities similar to those of classical radio halos and are hosted in clusters that are more common than the very massive systems hosting classical radio halos. Consequently, their observed rarity suggests that they could be intrinsically rare. For this reason in our model we do not attempt to take into account these sources; future surveys (LOFAR, EMU/WODAN) are necessary to get a firm conclusion on the occurrence of these sources in clusters.


\begin{acknowledgements}
We thank the referee for useful comments. RC and GB acknowledge partial support by PRIN- INAF2009 and ASI-INAF I/088/06/0.
\end{acknowledgements}

\end{document}